\documentclass{aa}
\usepackage{txfonts}
\usepackage{ifpdf}
\ifpdf
\usepackage[pdftex]{epsfig}
\else
\usepackage[dvips]{epsfig}
\fi
\usepackage{graphicx}
\usepackage{natbib}
\usepackage{color}
\usepackage{lscape}

\begin{document}
\bibpunct{(}{)}{;}{a}{}{,} 
\titlerunning{AGN in the zCOSMOS survey}
\title{Investigating the relationship between AGN activity and stellar mass in zCOSMOS galaxies at $0<z<1$ using emission line diagnostic diagrams.}

\author{M. Vitale\inst{1,2,3}
  \and M. Mignoli\inst{4}
  \and A. Cimatti\inst{3}
  \and S. J. Lilly\inst{5}
  \and C. M. Carollo\inst{5}
  \and T. Contini\inst{6,7}
  \and J.-P. Kneib\inst{8}
  \and O. Le Fevre\inst{8}
  \and V. Mainieri\inst{9}
  \and A. Renzini\inst{10}
  \and M. Scodeggio\inst{11}
  \and G. Zamorani\inst{4}
  \and S. Bardelli \inst{4}
  \and L. Barnes\inst{5}
  \and M. Bolzonella \inst{4} 
  \and A. Bongiorno \inst{12} 
  \and R. Bordoloi \inst{5} 
  \and T. J. Bschorr \inst{5} 
  \and A. Cappi  \inst{4}
  \and K. Caputi \inst{22} 
  \and G. Coppa  \inst{12}
  \and O. Cucciati  \inst{13}
  \and S. de la Torre  \inst{14}
  \and L. de Ravel \inst{14} 
  \and P. Franzetti  \inst{11}
  \and B. Garilli  \inst{11}
  \and A. Iovino  \inst{13}
  \and P. Kampczyk  \inst{5}
  \and C. Knobel  \inst{5}
  \and A. M. Koekemoer \inst{17} 
  \and K. Kovaˇc  \inst{5,19}
  \and F. Lamareille  \inst{6,7}
  \and J.-F. Le Borgne \inst{6,7} 
  \and V. Le Brun  \inst{8}
  \and C. L´opez-Sanjuan \inst{8} 
  \and C. Maier \inst{5,21}
  \and H. J. McCracken  \inst{19}
  \and M. Moresco \inst{3} 
  \and P. Nair  \inst{4}
  \and P. A. Oesch \inst{5,23} 
  \and R. Pello  \inst{6,7}
  \and Y. Peng  \inst{5}
  \and E. Perez Montero \inst{6,7,15} 
  \and L. Pozzetti \inst{4} 
  \and V. Presotto  \inst{13}
  \and J. Silverman  \inst{16}
  \and M. Tanaka  \inst{16}
  \and L. Tasca  \inst{8}
  \and L. Tresse \inst{8}
  \and D. Vergani  \inst{24}
  \and N. Welikala \inst{21} 
  \and E. Zucca  \inst{4}
}
 

\offprints{Mariangela Vitale, \email{vitale@ph1.uni-koeln.de}}

\institute{I. Physikalisches Institut, Universit\"at zu K\"oln, Z\"ulpicher Strasse 77, 50937 K\"oln, Germany
  \and Max-Planck Instutut f\"ur Radioastronomie, Auf dem H\"ugel 69, 53121 Bonn, Germany
  \and Dipartimento di Astronomia, Universit\`{a} di Bologna, via Ranzani 1, 40127 Bologna, Italy
  \and INAF - Osservatorio Astronomico di Bologna, via Ranzani 1, 40127 Bologna, Italy  
  \and Institute for Astronomy, ETH Zurich, 8093 Zurich, Switzerland
  \and Institut de Recherche en Astrophysique et Plan{\'e}tologie, CNRS, 14 Avenue Edouard Belin, F-31400 Toulouse, France
  \and IRAP, Universit{\'e} de Toulouse, UPS-OMP, Toulouse, France
  \and Aix Marseille Universit{\'e}, CNRS, LAM (Laboratoire d'Astrophysique de Marseille) UMR 7326, 13388, Marseille, France
  \and European Southern Observatory, Garching, Germany
  \and INAF-Osservatorio Astronomico di Padova, Vicolo dell’Osservatorio 5, 35122 Padova, Italy
  \and INAF-IASF Milano, Milano, Italy
  \and Max Planck Institut f\"ur Extraterrestrische Physik, Garching, Germany
  \and INAF Osservatorio Astronomico di Brera, Milan, Italy
  \and Institute for Astronomy, University of Edinburgh, Royal Observatory, Edinburgh, EH93HJ, UK
  \and Instituto de Astrofisica de Andalucia, CSIC, Apartado de Correos 3004, 18080 Granada, Spain
  \and Institute for the Physics and Mathematics of the Universe (IPMU), University of Tokyo, Kashiwanoha 5-1-5, Kashiwa, Chiba 277-8568, Japan
  \and Space Telescope Science Institute, Baltimore, MD 21218, USA
  \and Institut d’Astrophysique de Paris, UMR7095 CNRS, Universit{\'e} Pierre \& Marie Curie, 75014 Paris, France
  \and Max Planck Institut f\"ur Astrophysik, Garching, Germany
  \and Insitut d'Astrophysique Spatiale, Batiment 121, Universit{\'e} Paris-Sud XI \& CNRS, 91405 Orsay Cedex, France
  \and Department of Astronomy, University of Vienna, Tuerkenschanzstrasse 17, 1180 Vienna, Austria
  \and Kapteyn Astronomical Institute, University of Groningen, P.O. Box 800, 9700 AV Groningen, The Netherlands
  \and UC Santa Cruz/UCO Lick Observatory, 1156 High Street, Santa Cruz, CA 95064, USA
  \and INAF-IASF Bologna, Via P. Gobetti 101, I-40129, Bologna, Italy
}                                                                                                 


\abstract
{Active Galactic Nuclei (AGN) are thought to play an important role in galaxy evolution. It has been suggested that AGN-feedback could be partly responsible for quenching star-formation in the hosts, leading to transition from the blue cloud to the red sequence. The transition seems to occur faster for the most massive galaxies, where traces of AGN activity have been already found at $z<0.1$. The correlation between AGN activity, aging of the stellar populations and stellar mass still needs to be fully understood, especially at high redshifts.}{Our aim is to investigate the link between AGN activity, star-formation and stellar mass of the host galaxy at $0<z<1$, looking for spectroscopic traces of AGN and aging of the host. This work provides an extension of the existing studies at $z<0.1$ and contributes to shed light on galaxy evolution at intermediate redshifts.}{We used the zCOSMOS 20k data to create a sample of galaxies at $z<1$. We divided the sample in several mass-redshift bins to obtain stacked galaxy spectra with an improved S/N. We exploited emission-line diagnostic diagrams to separate AGN from star-forming galaxies.}{We found indication of a role for the total galaxy stellar mass in leading galaxy classification. Stacked spectra show AGN signatures above the $\log M_*/M_{\odot}>10.2$ threshold. Moreover, the stellar populations of AGN hosts are found to be older with respect to star-forming and composites galaxies. This could be due to the the tendency of AGN to reside in massive hosts.}{The dependence of the AGN classification on the stellar mass is in agreement with what has been already found in previous researches. It is consistent with, together with the evidence of older stellar populations inhabiting the AGN-like galaxies, the downsizing scenario. In particular, our evidence point to an evolutionary scenario where the AGN-feedback is capable of quenching the star formation in the most massive galaxies. Therefore, the AGN-feedback is the best candidate for initiating the passive evolutionary phase of galaxies.}

\maketitle

\section{Introduction}
In recent years, a better overview of galaxy formation and evolution has been developing, with a great contribution coming from large cosmic surveys. In particular, the local galaxies colour function has been discovered to be a bimodal function \citep{Strateva2001,Kauffmann2003a,Baldry2004,Balogh2004,Hogg2004}. The two dominant sequences are associated with the ellipticals+S0s population (red distribution, or \textit{red sequence}) and the spirals+irregulars population (blue distribution, or \textit{blue cloud}), while galaxies showing intermediate properties are considered as composites or transitional objects (\textit{green valley}). Early-type galaxies (ETGs) dominate at the top-end of the local galaxy mass function where, on the contrary, late-type galaxies (LTGs) become progressively more rare. The discovery of how the black hole (BH) mass strongly correlates with the properties of the hosting galaxies \citep{Magorrian1998,Ferrarese2000}, has generated new hypotheses in linking the black hole growth with galaxy formation and evolution. The mass of the host seems to play a key role in the latter process as well. According to the ``downsizing'' scenario \citep{Cowie1996,Kodama2004,Thomas2005,Tanaka2005,Juneau2005,Bundy2006}, the sites of active star-formation include high-mass galaxies at high redshift ($z\gtrsim1$) but only intermediate- and low-mass galaxies at $z\lesssim1$. This scenario is supported by several studies, e.g. it has been found that luminous and massive spheroidals build up most of their stellar mass prior to $z\simeq1$, while low mass ellipticals continue to grow after this cosmic time \citep{Heavens2004,Treu2005,vanderWel2005,Jimenez2005}. Downsizing may be a natural expectation in a hierarchical galaxy formation scenario, provided that there is a mechanism that quenches star formation at earlier times in massive galaxies. As a support to downsizing and the existence of a link between mass and galaxy evolution, \citet{Bundy2004} found that the relative abundance of red galaxies with intermediate mass ($M\approx6\times10^{10}M_{\odot}$) increases by a factor of $3$ from $z\sim1.2$ to $z\sim0.55$, while the number of blue late-type galaxies declines significantly.\newline
The fast suppression of the star formation activity (\textit{quenching}) is currently thought to initiate passive galaxy evolution and to be largely responsible for the growing abundance of galaxies on the red sequence \citep{Faber2007}. The quenching mass ($M_Q$), namely the stellar mass above which galaxies start their passive phase, has found to be a function of the redshift \citep{Bundy2004}. Blue and star-forming galaxies at their quenching mass are redder than the average blue population, suggesting a transion from blue- to red-type galaxies.\newline
The star-formation quenching could be explained by several mechanisms. The observed decline of the galaxy star formation from $z\sim1$ to $z=0$ can be due to a gradual \textit{gas exhaustion} \citep{Cooper2008}. Another mechanism that could explain the quenching is \textit{major mergers}. The interaction between a galaxy and its neighbours could indeed produce gravitational torques on the gas, reducing its angular momentum and sending it toward the galactic center after an initial starburst phase \citep{Martig2009}. As a result, fuel is available for accretion, allowing the growth of the central Black Hole (BH) rather than the formation of new stars in the disk. However, it has been established that mergers cannot be the main cause to the build up of the galaxy bimodality, since luminous and massive old galaxies formed throughout this process were already common at $z\sim1$ \citep{Conselice2007} and their number density only declines above this value. If merger events were efficient in forming galaxies at $z<1$, we would expect an increasing number of massive objects at low redshift, which has not been observed. Similarly, a role for active galactic nuclei (AGN) has been suggested \citep{Ciotti1997,Binney2004,Silk2005,Springel2005a}. BHs are thought to be a basic constituent of most massive systems \citep{Richstone1998} and they have also been identified in some late-type and dwarf galaxies \citep{Filippenko2003}. During the BH accretion, AGN release a large amount of energy, which could be partially absorbed by the host galaxy to quench the star formation and eventually transform blue galaxies into red galaxies \citep[e.g.][]{Silk1998,Croton2006,Narayanan2008}. According to some calculations \citep{Cattaneo2009}, even a small fraction ($<1\%$) of the energy released within a bulge during the BH accretion would be sufficient to heat and/or blow away the entire gas content. Consistently with this framework, \citet{Hopkins2006a} further developed \citet{Sanders1988b} scenario in which starburst, quasar active phase, BHs growth and elliptical galaxies are connected to each other in a galaxy evolutionary sequence and the gas consumption is coupled with supernova-driven winds and/or AGN-feedback to quench the star formation.\newline
The hypothesis of the AGN-feedback is strengthened by the fact that a large amount of galaxies have been found to go through an active phase. Summed over all Hubble types, roughly half of all galaxies can be considered as active \citep{Ho1997b}, even though some of them show a fairly low activity level. The amount of active galaxies becomes more remarkable for galaxies with a prominent bulge component, rising to $50-70\%$ for Hubble types E-Sb, whereas the detection rate of AGN drops towards later Hubble types (Sc and later), where about $80\%$ of the galaxies host a star-forming nucleus instead. This suggests a clear dependence of nuclear activity on the Hubble type \citep{Kauffmann2003a,Miller2003}. The observed distribution of Hubble types for galaxies hosting active versus inactive nuclei leads to the expectation that the blue and red populations could have not only different nuclear properties but global as well. In particular, it seems that the observed high fraction of AGN in the green valley points to the connection between the end of the star formation phase, observed in a change of the galaxies color, and the rise of the AGN activity \citep{Silverman2009,Schawinski2010}.\newline
AGN can be identified by a variety of methods: an unusually blue continuum, strong radio or X-ray emission, the presence of strong or broad emission-lines. In nearby galaxies hosting low-luminosity AGN, we expect the non-stellar signal of the nucleus to be weak with respect to the one coming from the host galaxy, thus one of the best and least biased methods to search for AGN is to conduct a spectroscopic survey of a complete, optical-flux limited sample of galaxies and study their spectral properties. The line-emitting gas in AGN and star-forming galaxies (SFGs) is powered by two different ionizing mechanisms, producing different emission line ratios: accretion around black hole and photoionization by hot massive OB stars. It is important to note that emission lines in narrow-line AGN have a considerably great ionization range and that low-ionization lines are stronger than in normal star-forming galaxies. It has been demonstrated that ordinary O-type stars do not produce sufficiently strong low-ionization lines, accounting for the observed spectra, as they only account for a few percentage of the total blue light at most \citep{Sarzi2005}. Therefore, photoionization arising from powerful central non-thermal source is the ideal candidate to explain the excitation mechanism in galaxies which display strong low-ionization emission-lines \citep{Ho2005}. \newline
Emission-line diagnostic diagrams \citep{Baldwin1981,Veilleux1987,Kewley2003,Lamareille2004,Lamareille2010} represent a powerful tool for probing the nature of the dominant ionizing source in galaxies, hence to distinguish between objects dominated by star formation and galaxies where the nuclear activity is more relevant. Narrow-line AGN can be identified in spectroscopic galaxy surveys by the ratio of some distinctive emission-lines, such as [NII]$\lambda6583$ \AA\ (hereafter [NII]) or [SII]$\lambda\lambda6717,31$ \AA\ (hereafter [SII], referred to the combined luminosity of the doublet) over H$\alpha\lambda6563$ \AA\ and [OIII] $\lambda5007$ \AA\ over H$\beta\lambda4861$ \AA{} (hereafter H$\alpha$, [OIII] and H$\beta$, respectively). The lines that appear in the reddest part of the spectrum are shifted out of the visible wavelength range for $z>0.5$. In this case, it is possible to use the [OII]$\lambda\lambda3726+3729$ \AA\ line (hereafter [OII]) instead \citep{Rola1997,Lamareille2004,Perez-Montero2007}, which is optically visible up to $z\sim1$. \newline
The aim of this work is to investigate whether it is possible to highlight some evidence of AGN activity in suppressing star-formation, with a particular look on a possible trend of the galaxy total stellar mass. This attempt has been already made in the past at low-redshift ($z<0.1$) and the analysis is now extended to redshift up to $z\sim1$. \citet{Kauffmann2003a} find, by using Sloan Digital Sky Survey (SDSS) data, that galaxies divide in two distinct families at a stellar mass threshold of $3\times10^{10}~M_{\odot}$. The least massive galaxies show young stellar populations and the low concentrations which are typical of LTGs. As the stellar mass increases, galaxies show older stellar populations and higher concentrations, typical of ETGs. By using emission-line diagnostic diagrams, they show a dependence of the AGN-detection rate on mass, although the authors warn that many objects are classified as ``composites'' (contributions to the emission-line spectrum come by both star-formation and AGN) because of the large size of the SDSS fiber, which can collect up to $40$\% of the total light from the galaxy. Blue star-forming objects have been found to have masses at the low end of the mass function, while the majority of the ETGs lie at the high-end of the mass function \citep{Baldry2004,Schawinski2007}. Very recently, however, \citet{Aird2012} argued that the finding of more AGN in massive hosts is due to selection effects. The latter seem to be driven by the Eddington-ratio distribution of AGN in galaxies of any stellar mass. In particular, AGN are more easily detected in massive hosts as sources accreting at low Eddigton rate are more luminous than in less massive hosts.\newline
We analyze the evolution of galaxy spectral properties up to $z\sim 1$ by using galaxies from the zCOSMOS-Bright 20k sample \citep{Lilly2007}. With the large number ($\sim2\times 10^4$) of observed objects, we are able to measure emission lines from high S/N stacked spectra. The paper is organized as follows. In \textit{Section 2} we present the zCOSMOS survey and the sample selection. In \textit{Section 3} we present the spectral analysis of the stacked, stellar-continuum subtracted galaxy spectra. In \textit{Section 4} we exploit a set of classical and more recent diagnostic diagrams to identify the main ionizing process inside galaxies and look for a trend of the stellar mass. \textit{Section 5} contains our results and discussion. Main findings and conclusions are presented in \textit{Section 6}.

\section{Data}
\subsection{COSMOS and zCOSMOS surveys}
The COSMOS survey is a large HST-ACS survey, with $I$-band exposures down to $I_{AB}=28$ on a field of $2$ deg$^2$ \citep{Scoville2007}. The COSMOS field has been the object of extensive multiwavelength ground- and space-based observations spanning the entire spectrum: X-ray, UV, optical/NIR, mid-infrared, mm/submillimeter and radio, providing fluxes measured over $30$ bands \citep{Hasinger2007,Taniguchi2007,Capak2007,Lilly2007,Sanders2007,Bertoldi2007,Schinnerer2007,Koekemoer2007,McCracken2010}.\newline
The zCOSMOS survey was planned to provide high-quality redshift information to the COSMOS field \citep{Lilly2007}. It benefitted of $\sim600$ h of observations at VLT using the VIMOS spectrograph and it consists of two parts: zCOSMOS-bright and zCOSMOS-deep. The zCOSMOS-deep targets $\sim10\ 000$ galaxies within the central $1$ deg$^2$ of the COSMOS field, selected through color criteria to have $1.4\lesssim z\lesssim3.0$. The zCOSMOS-bright is purely magnitude-limited and covers the whole area of $1.7$ deg$^2$ of the COSMOS field. It provides redshifts for $\sim20\ 000$ galaxies down to $I_{AB}\lesssim 22.5$ as measured
from the HST-ACS imaging. The success rate in redshift measurements is very high, $95$\% in the redshift range $0.5<z<0.8$, and the velocity accuracy is $\sim100$ km s$^{-1}$ \citep{Lilly2009}. Each observed object has been assigned a flag according to the reliability of its measured redshift. Classes $3.x$, $4.x$ redshifts, plus Classes $1.5$, $2.4$, $2.5$, $9.3$, and $9.5$ are considered a secure set, with an overall reliability of 99\% \citep[see][for details]{Lilly2009}.\newline
Our work is based on the the zCOSMOS-bright survey final release: the so called $20k$ sample, totaling
$16\ 623$ galaxies with $z\lesssim2$ and secure redshifts according to the above flag classification. It includes $18\ 206$ objects in total, including stars.\newline
For objects brighter than $I_{AB}=22.5$ and without secure spectroscopic redshift, photometric data from the COSMOS survey provides good quality photometric redshifts \citep{Ilbert2009}. Based on a comparison with the zCOSMOS spectroscopic redshifts, \citet{Ilbert2009} estimate an accuracy of $\sigma_{zphot}=0.007\times(1+z_s)$ for galaxies brighter than $I_{AB}=22.5$.\newline
For all galaxies brighter than $I_{AB}=22.5$, absolute rest-frame magnitudes and stellar masses were obtained using standard multi-color spectral energy distribution (SED) fitting techniques, using the secure spectroscopic redshift, if available, or the photometric one. Stellar masses were obtained using the $hyperzmass$ code \citep{Pozzetti2010,Bolzonella2010}, by assuming a Chabrier initial mass function \citep{Chabrier2003}.
To establish mass completeness, \citet{Pozzetti2010} define at each redshift a minimum mass ($M_{min}$) above which the derived galaxy stellar mass function is essentially complete (all types of galaxies are potentially observable). To derive $M_{min}$, the limiting stellar mass ($M_{lim}$) of each galaxy is calculated, namely the mass the galaxy would have, at its spectroscopic redshift, if its apparent magnitude were equal to the limiting magnitude of the survey. Then, in order to derive a representative limit for the sample, the authors use the $M_{lim}$ of the $20$\% faintest galaxies at each redshift. $M_{min}(z)$ is defined as the upper limit of the $M_{lim}$ distribution below which $95$\% of the $M_{lim}$ values at each redshift are represented. This is assumed to be the completeness limit of the galaxy stellar mass function.

\subsection{Analyzed galaxy sample}
In order to create the bins that collect the galaxies to be stacked, we used the Bright zCOSMOS spectroscopic catalogue (\textit{v4.12} version). In case of AGN that are more luminous than the host galaxy, as the type-1 AGN where the emission is not screened by the dusty torus \citep[AGN-unification model by ][]{Padovani1992}, we are not able to see the emission lines from the host galaxy. Therefore, we remove type-1 AGN to avoid overshining issues. This was performed by excluding all the objects with broad-emission lines. Furthermore, we decided to exclude from our sample all the sources that do not bring an indication of well measured spectroscopic redshift. These latter classes and the broad-line AGN constitute, all together, around $6\%$ of the original sample within $0.1<z<1$ ($16\ 678$ objects). In particular, we exclude $892$ low-flag objects and $35$ broad-line AGN. The galaxies with reliable flag are then $15\ 715$. Another significant fraction of galaxies is lost during the binning process due to exclusion of underrepresented regions of the redshift-stellar mass diagram (see next section). After applying these selection criteria, we are able to study roughly half of the original zCOSMOS-Bright sample.

\section{Spectral analysis}
In this chapter we present how we exploited the large number of zCOSMOS galaxies by dividing them in various redshift and mass bins. This was done in order to obtain average spectra with high S/N and to investigate the evolution of the spectral properties of the stacked galaxies. We aim to study the variation of average properties of galaxies that are representative of a particular mass-redshift bin. After obtaining the spectra, we estimate and subtract the stellar contribution. Then, we measure line fluxes and compute the visual extinction. Finally, we use diagnostic diagrams to distinguish between the main ionizing mechanism inside these objects. Several programs and tasks have been used for this work, most of them running in the \textit{Image Reduction and Analysis Facility} (IRAF) environment.
 \begin{figure*} 
  \centering
  \includegraphics[width=16cm,angle=0]{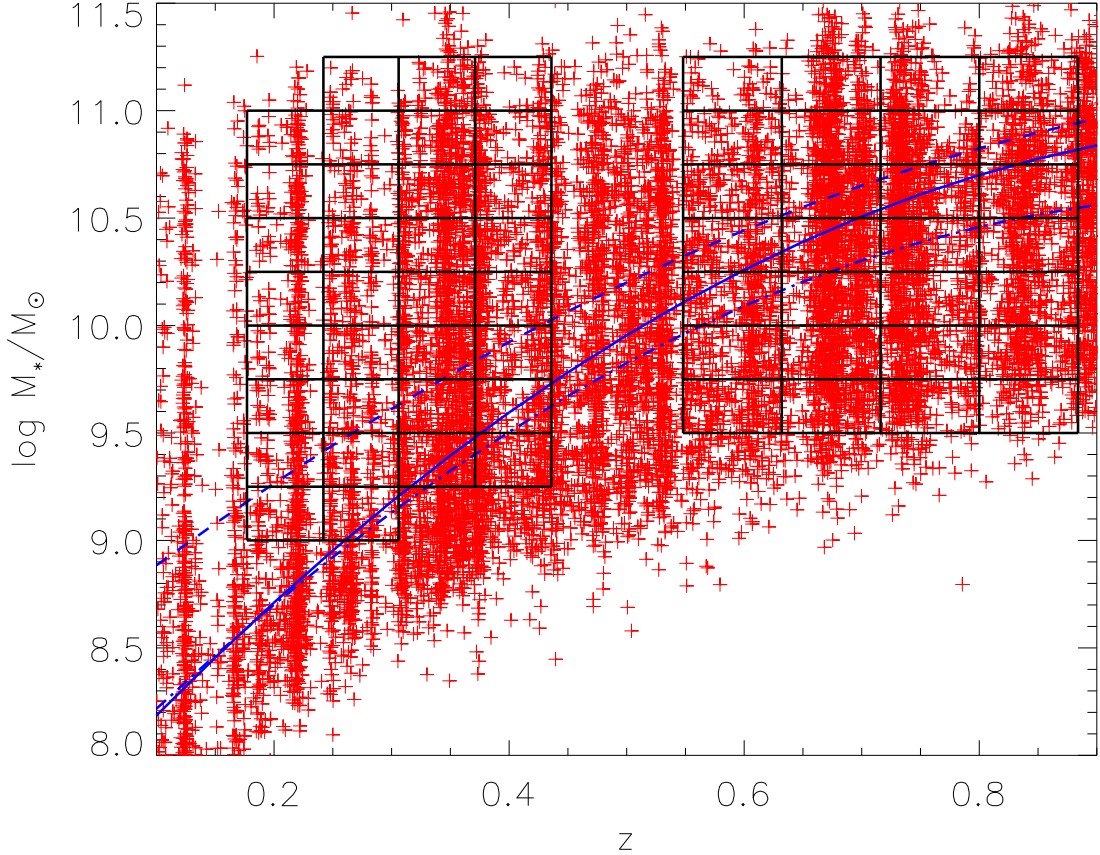}
  \caption{\label{binning} The stellar mass of zCOSMOS 20k galaxies selected for our study (secure flags, no broad-line AGN) plotted as a function of redshift. The blue dashed lines represent, respectively, the mass completeness limits of the global galaxy population (solid), the population of early-type (dashed) and the late-type galaxies (dot-dashed). The black boxes mark the $61$ mass-redshift bins that define the stacked galaxies we study with the diagnostic diagrams.}
 \end{figure*}

\subsection{Sample binning}
In order to search for variations of galaxy spectral properties as a function of redshift and stellar mass, we divided our large sample in bins. The bin width was tested in order to achieve a good signal to noise with the largest number of mass-redshift bins. In particular, we tested the increase in S/N according to the increase in the number of stacked spectra. The first indication of a link between total galaxy stellar mass and classification in the spectral types - AGN, composite galaxies, star-forming galaxies - made us decide for a fine grid in $M_*$. On the contrary, the variations we applied to the size of the redshift intervals did not affect the classification significantly. We afterwards decided for the number of $z$ bins to be equal to $8$, with a redshift step $\sim0.1$ dex. We divided the sample in $9$ mass bins, with a mass step of $0.25$ dex. The bins include few hundreds of objects for each redshift and mass range (See Tab. \ref{statistics}). For our sample selection, mass completeness is achieved when including in the bins galaxies above the completeness curves \citep{Pozzetti2010}. While at lower redshift most bins lie above the completeness limits, at higher redshifts roughly half of the bins are placed below the confidence curves. Although, in this case, the sample suffers from incompleteness, we chose to study all the high-redshift bins in order to probe a comparable dynamical range of masses to the low-redshift case. \newline
The choice of the bins comes from the following further considerations:
\begin{itemize}
 \item below $9$ and over $11.25$ $\log M_*/M_{\odot}$, objects are poorly represented in our sample. The S/N ratio of the average spectra made out of the stacking of an insufficient number of galaxies would not allow precise line measurements;
 \item 8 bins (4 per region covered by each of the two types of diagnostic diagrams, $z\lesssim0.5$ for the low-redshift diagnostic diagrams and $z\gtrsim0.5$ for the high-redshift diagnostic diagram) are a good trade off between the need to explore the cosmological evolution of the galaxy spectral properties and to keep the number of objects per bin (thus the S/N) high enough.
\end{itemize}
 \begin{table*} 
  \caption{\label{statistics} Number of galaxies in each mass-redshift bin of the sample.}
  \centering
    \begin{tabular}{|c|cccc|cccc|}
    \hline
     $M_m$ / $z_m$ &$0.21$&$0.27$&$0.34$&$0.4$&$0.59$&$0.67$&$0.76$&$0.84$\\
    \hline
    \hline
     $11.12$&-&24&83&47&52&133&101&114\\
     $10.87$&51&42&148&107&106&241&237&189\\
     $10.62$&63&45&196&118&146&324&296&221\\
     $10.37$&68&69&177&143&168&323&272&198\\
     $10.12$&57&66&189&156&187&283&216&173\\
     $9.87$&52&79&214&152&179&286&219&201\\
     $9.62$&76&73&213&162&193&238&185&87\\
     $9.37$&83&95&266&211&-&-&-&-\\
     $9.12$&108&109&-&-&-&-&-&-\\
    \hline
    \end{tabular}
   \tablefoot{ The table reflects the order of the mass-redshift bins shown in Fig. 1. The median mass (first column) is expressed in $\log M_*/M_{\odot}$. The median redshift is indicated on top of each row.}
 \end{table*} 
For diagnostic diagrams based on [NII], [SII], H$\alpha$, [OIII] and H$\beta$, the affected rest-frame spectral range is from $4800$ to $6650$ \AA{}. The zCOSMOS spectra cover the $5650$ to $9550$ \AA{} range, after cutting off $100$ \AA{} at both extremes to avoid noisy regions where the flux calibration is more uncertain. The useful redshift range for the low-redshift diagnostic diagrams is then $0.177\leq z\leq 0.436$. In the same way, the redshift range for the high-redshift diagnostic diagram is: $0.548\leq z\leq 0.884$.\newline
In Fig. \ref{binning} we plot the zCOSMOS galaxies as a function of the total galaxy stellar mass. The plot displays our bin selection and the large scale structure within the zCOSMOS field. The general trend is a clear increase of the average stellar mass with the redshift due to the well-known selection effects of a magnitude-limited survey, with overdense regions (filaments), e.g. at $0.3\gtrsim z\gtrsim 0.4$ or $0.7\gtrsim z\gtrsim 0.8$, and underdense regions (voids), e.g. at $z\sim 0.25$. At low redshift the minimum mass ($M_{min}$) of the global galaxy population is better represented by the blue population, whereas it is closer to the red population if we are at higher redshift. This is because blue galaxies dominate the mass function at low redshift, due to selection effects. On the other hand, at high redshift the global $M_{min}$ shifts towards the limit of the reddest population, since the galaxies have a higher mass-to-light ratio ($M/L$) on average.

\subsection{Stacking}
After the binning process, the spectra were stacked. Since each stacked spectra is the result of the average of up to several hundreds of single galaxy spectra included in a single bin, the gain in S/N is considerable and allows a more accurate line fitting. In order to create the composites, each spectrum was shifted to the rest-frame. The spectra were normalized in wavelength ranges always present in the observed spectroscopic window and lacking prominent spectral features. The wavelengths of the chosen ranges are: $5300-5800$ \AA{} for $0.177<z<0.371$; $4500-4800$ \AA{} for $0.371<z<0.716$; and $3400-3700$ \AA{} for $0.716<z<0.884$. Stacked spectra are shown in Appendix A. The main emission- and absorption-lines are labelled on top of each figure.\newline
The stacked zCOSMOS spectra present, especially for the objects with low masses, a quite blue and featureless continuum at low redshift. The spectra are characterized by the presence of strong low-ionization emission lines such as [NII], [SII] and [OIII] and strong recombination lines of the Balmer series, as expected from the spectral contribution of a young stellar population. On the other hand, the continuum is much redder and displays a strong D4000 break at higher redshift, which is indicative of an older stellar population. The spectra show that the most massive galaxies are spectroscopically more evolved at at a given redshift, according to the downsizing scenario.

\subsection{Stellar continuum subtraction}
Galaxy emission lines - in particular the ones from the Balmer series - are often contaminated by absorption lines produced by the atmospheres of fairly massive stars within the galaxies. Since our spectra are characterized by both emission from the hot ionized gas and absorption from the stellar component, we need an effective strategy to make precise measurements of spectral features involved in the diagnostic diagrams.
The technique we adopted to remove the starlight from an integrated spectrum is to subtract a suitable model of stellar continuum to each stacked galaxy spectrum to obtain an almost pure emission-line spectrum. This is particularly helpful in correcting the strength of Balmer lines, which are involved in both the AGN identification and the computation of the visual extinction. The procedure also aims to additional information on stellar populations, like age and metallicity, that would have been lost if we had focused on emission lines only.
 \begin{figure*} 
  \centering
  \includegraphics[width=14cm,angle=0]{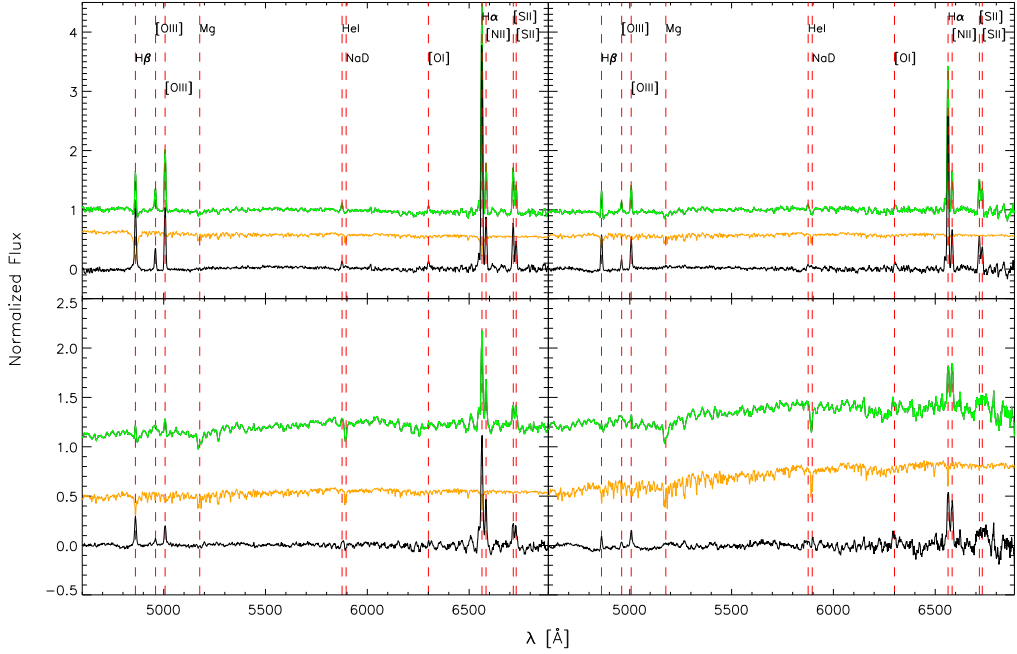}
  \caption {Four examples of stellar continuum subtraction at low redshift ($z<0.436$, one per redshift bin). For each panel, the original stacked spectrum (green), the most suitable stellar continuum to be subtracted (orange) and the resulting zCOSMOS spectrum after the stellar continuum subtraction (black) are shown. The main emission and absorpion lines are labelled on top. From top to bottom and from left to right both total stellar mass and redshift increase. Top-left panel: stacked spectrum of the bin $9.25\leq \log M_*/M_{\odot}<9.5$, $0.177 \leq z<0.242$, stellar template representing a $290$ Myr old simple stellar population (SSP) with Z=0.2 and E(B-V)=0.4; top-right panel: $9.75< \log M_*/M_{\odot}\leq 10$, $0.242 \leq z<0.306$, $12$ Gyr old stellar template with exponential decay of the star formation, Z=0.2 and E(B-V)=0.2; bottom-left panel: $10.25\leq \log M_*/M_{\odot}<10.5$, $0.306 \leq z<0.371$, $900$ Myr old SSP with Z=0.5 and E(B-V)=0.4; bottom-right panel: $10.75\leq \log M_*/M_{\odot}<11$, $0.371 \leq z<0.436$, $11$ Gyr old SSP with Z=0.2 and E(B-V)=0.4. 
 The stacked spectra before stellar continuum subtraction and the stellar templates have been shifted up of arbitrary quantities to offer a clearer view.}
 \label{cont_subtracted}
 \end{figure*} 
 \begin{figure*} 
  \centering
  \includegraphics[width=14cm,angle=0]{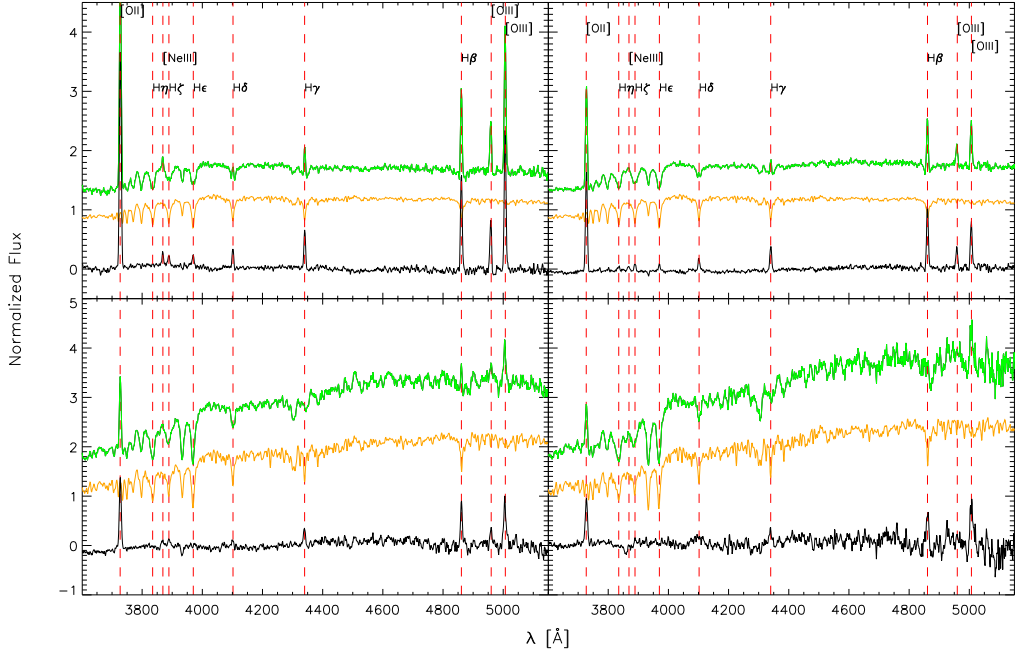}
  \caption {As in Fig. \ref{cont_subtracted}, but at higher redshift ($z>0.548$). Top-left panel: stacked spectrum of the bin $10.5\leq \log M_*/M_{\odot}<10.75$, $0.548 \leq z<0.632$, stellar template representing a $1.4$ Gyr old SSP with Z=0.08 and E(B-V)=0.2; top-right panel: $10.75< \log M_*/M_{\odot}\leq 11$, $0.632 \leq z<0.716$, $1.4$ Gyr old SSP with Z=0.2 and E(B-V)=0.2; bottom-left panel: $10.75\leq \log M_*/M_{\odot}<11$, $0.716 \leq z<0.8$, $1.4$ Gyr old SSP with Z=0.08 and E(B-V)=0.2; bottom-right panel: $11\leq \log M_*/M_{\odot}<11.25$, $0.8 \leq z<0.884$,  $1.4$ Gyr old SSP with Z=0.2 and E(B-V)=0.2. The stacked spectra before stellar continuum subtraction and the stellar templates have been shifted up of arbitrary quantities to offer a clearer view.}
 \label{cont_subtracted2}
 \end{figure*} 
For this purpose, we have developed an user-written IRAF task that performs the spectra continuum subtraction. The spectral library to be fitted to the stacked spectra must contain enough information on various absorption features, in order to be able to simulate the stellar components of galaxies within a wide range of ages and metallicities. We adopted the \textit{Bruzual and Charlot (BC) library}, which is a library of stellar population synthesis models computed by Bruzual and Charlot using their \textit{Isochrone Synthesis Spectral Evolutionary Code} \citep[][hereafter BC03]{BruzualCharlot2003}. This code allows to predict the spectral evolution of stellar populations in various ranges of ages and metallicities at a resolution of $3$ \AA{}, across the whole wavelength range (from $3200$ \AA{} to $9500$ \AA{}). Models with different time scales of star formation, initial mass functions and metallicities well reproduce the spectral and photometric properties of nearby galaxies with various morphological types, from young irregulars to elliptical galaxies. For each BC template, we created $5$ spectra with increasing B(E-V) values ($0.2$, $0.4$, $0.6$, $0.8$ and $1$) in order to take reddening due to dust into account. A total of $234$ stellar templates were compared to the stacked spectra to find the most suitable to be subtracted to each composite. The continuum subtraction task degrades the resolution of the template spectra to the one of zCOSMOS (R$\sim 600$). Then, the template is chosen to have the lowest residuals in pre-selected regions of the continuum subtracted sample which are free from strong emission lines. These spectral windows have a typical width of $\sim 100$ \AA. They span the range between $3600$ and $7300$ \AA, for an overall coverage of about $2300$ \AA.\newline
The stellar continuum subtraction procedure successfully corrects for the underlying stellar absorption (Fig. \ref{cont_subtracted}, \ref{cont_subtracted2}). This feature is particularly noticeable in the highest redshift bins (Fig. \ref{cont_subtracted2}), where the strong stellar contribution to the galaxies spectra comes from older stellar populations.

\subsection{Emission line measurement}
We measured the flux of [OII], H$\gamma\lambda4340$ (hereafter H$\gamma$), H$\beta$, [OIII], H$\alpha$, [NII] and [SII]. We used the \textit{Splot} interactive IRAF task to make spectral measurements of gaussian-like emission lines. H$\alpha$ and H$\beta$, along with the oxygen lines, [NII] and [SII], are used in the exploitation of the diagnostic diagrams. The errors on the measurements of the emission lines in our stacked spectra corrected for the stellar absorption are typically less than $20\%$, estimated on the basis of repeated independent measurements. The errors mainly depend on the line intensity and whether the emission line is blended with other spectral features. The resolution of the zCOSMOS spectra is sufficient to resolve the [SII] doublet in most of the composites, but it is not sufficient to resolve the [OII] doublet, which always appears as a blended feature.\newline
For the highest stellar masses, the emission line measurements are more difficult because the S/N ratio is lower due to poorer statistics, besides the fact that the lines become weaker and the continuum stronger, especially at high redshift.

\subsection{Reddening correction}
Observed fluxes need to be corrected for extinction. The total visual extinction, $A_V$, can be determined by using the ratio between two recombination lines of the Balmer series. For this purpose, H$\alpha$ and H$\beta$ are commonly preferred as they are strong lines placed in an easily accessible spectroscopic position. In this work, we assumed the H$\alpha$/H$\beta$ theoretical ratio to be $2.86$ and the H$\gamma$/H$\delta$ to be $0.47$ \citep[][case B recombination, temperature T$=10^4$ K, electron density n$_e=10^2$ cm$^{-2}$]{Osterbrock1989} and we assumed the Calzetti extinction curve \citep{Calzetti1994}, with $R_V=4.05\pm0.8$.\newline
In diagnostic diagrams that make use of lines which are close to each other in the spectrum, their ratio is relatively insensitive to reddening effects. Nevertheless, we chose to correct for the visual extinction because the [OII] and H$\beta$ lines are quite far away from each other in wavelength, thus their ratio is sensitive to reddening.\newline

\section{Diagnostic diagrams}
In this work, we adopted the following emission-line diagnostic diagrams:
\begin{itemize}
 \item {[NII]/H$\alpha$ versus [OIII]/H$\beta$ ($z<0.436$)} \citep{Baldwin1981,Veilleux1987,Kewley2001,Kauffmann2003a,Kewley2006}
 \item {[OII]/H$\beta$ versus [OIII]/H$\beta$ ($z>0.548$)} \citep{Tresse1996,Rola1997,Lamareille2004,Lamareille2010}
 \item MEx diagram \citep{Juneau2011}
\end{itemize}
The redshift range $0.436<z<0.548$ has been excluded from our study because [NII] and H$\alpha$ are redshifted outside the optical spectra and [OII] is not yet visible in the blue part of the spectral range. Therefore, it is not possible to use the first two diagnostic diagrams. In case of the MEx diagram, we decided to keep the same grid of stellar mass and redshift bins. Although it would be technically possible to include the missing redshift range when using the MEx diagram, we expect that this would not change our finding that the AGN classification is more dependent on the stellar mass rather than on the redshift.

\subsection{Low redshift diagnostic diagrams ($0.177<z<0.436$)}
  \begin{figure*} 
   \includegraphics[width=18cm,angle=0]{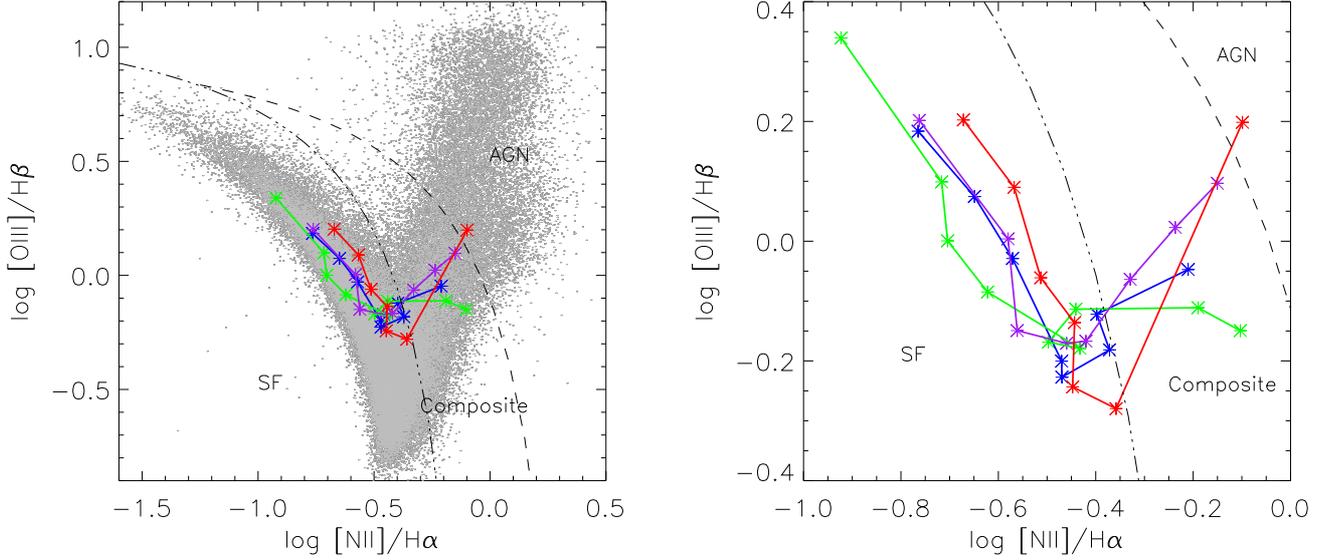}
   \caption{Classical [NII] diagnostic diagram. In both panels, a theoretical demarcation curve (dashed, Kewley et al. 2001), separates star-forming galaxies and composites from AGN, while an empirical demarcation curve (three-dot-dashed, Kauffmann et al. 2003) separates pure SFGs from composites and AGN. Each color represents the trend with stellar mass for a fixed redshift bin: $0.177<z<0.242$ (blue), $0.242<z<0.306$ (green), $0.306<z<0.371$ (purple) and $0.371<z<0.436$ (red). The total stellar mass increases from left to right. In the left panel, the SDSS galaxies from \citet{Vitale2012} are plotted in gray. The right panel offers a a closer view of the zCOSMOS galaxies.}
  \label{nii}
  \end{figure*} 
In the [NII]/H$\alpha$ versus [OIII]/H$\beta$ or Baldwin-Phillips-Terlevich (BPT) diagram \citep{Baldwin1981}, galaxies are distributed in two arms. The increase of the [NII]/H$\alpha$ ratio is a linear function of the nebular metallicity and it presents a saturation point (Denicol, Terlevich $\&$ Terlevich 2002; Kewley $\&$ Dopita 2002, Pettini $\&$ Pagel 2004) above which any further increase in the [NII]/H$\alpha$ value is only due to AGN contribution \citep{Kewley2006,Stasinska2006}. The BPT emission-line diagnostic diagram makes use of different demarcation curves, both theoretical and based on observations. The first curve was derived theoretically by Kewley et al. (2001) in order to find an upper limit for star-forming galaxies (Fig. \ref{nii}). \citet{Kauffmann2003a}, using the large sample of emission line galaxies in the SDSS, defined a demarcation which traces more closely the observed lower left branch attributed to purely star forming galaxies. This results in a larger fraction of galaxies residing on the AGN side (Fig. \ref{nii}). According to the latter classification, some of the AGN and composite galaxies with widely spread distributions of metallicity and ionization parameters are degenerate with star-forming galaxies. This is especially the case, for example, of the [SII]/H$\alpha$ versus [OIII]/H$\beta$ diagram.\newline
Objects that are classified as AGN in the [NII]-based diagram can be classified as star-forming galaxies in the [SII]/H$\alpha$ versus [OIII]/H$\beta$  diagram. This is thought to be related to the enhancement of [SII] lines in starburst galaxies, due to the mechanical energy released into the gas by supernovae and stellar winds \citep[also called shock excitation,][]{Dopita2002}. However, it is still not clear whether the [SII] enhancement is driven by highly ionizing photons produced by accretion onto the super massive black hole or if the SNe winds are a sufficiently powerful source of ionization. The [SII]-based diagnostic diagram, due to the fringing and the lines blending, is more affected by errors and leads to a higher probability of misclassifying galaxies with respect to the [NII]-based diagram. For these reasons, the diagnostic diagram that makes use of the [SII] doublet is not used in this paper to draw conclusions but it is only shown later as a comparison.\newline
In Fig. \ref{nii} (left panel), we show the zCOSMOS data superimposed to the SDSS data from \citet{Vitale2012}. Each color represents a different redshift bin and each point along the curve represents a different mass-bin. The right panel offers a zoom on the region of the diagram where our data are placed. The stacked spectra show a progressive shift towards the AGN region - on the right-hand side of the diagram - for increasing stellar mass (from left to right). The objects with the highest mass are placed on the right of the Kauffmann demarcation curve, which separates the region where galaxies start to show the sign of significant nuclear activity. The stacks follow, for a fixed redshift bin, traks  that move from the upper part of the SF galaxies sequence, down to the metallicity sequence \citep{Kewley2006}. For $\log M_*/M_{\odot}>10.2$, the objects start being classified as composites or AGN, in agreement with what already suggested by \citet{Kauffmann2003a} for $z<0.1$.
 \begin{figure*} 
  \includegraphics [width=18cm,angle=0]{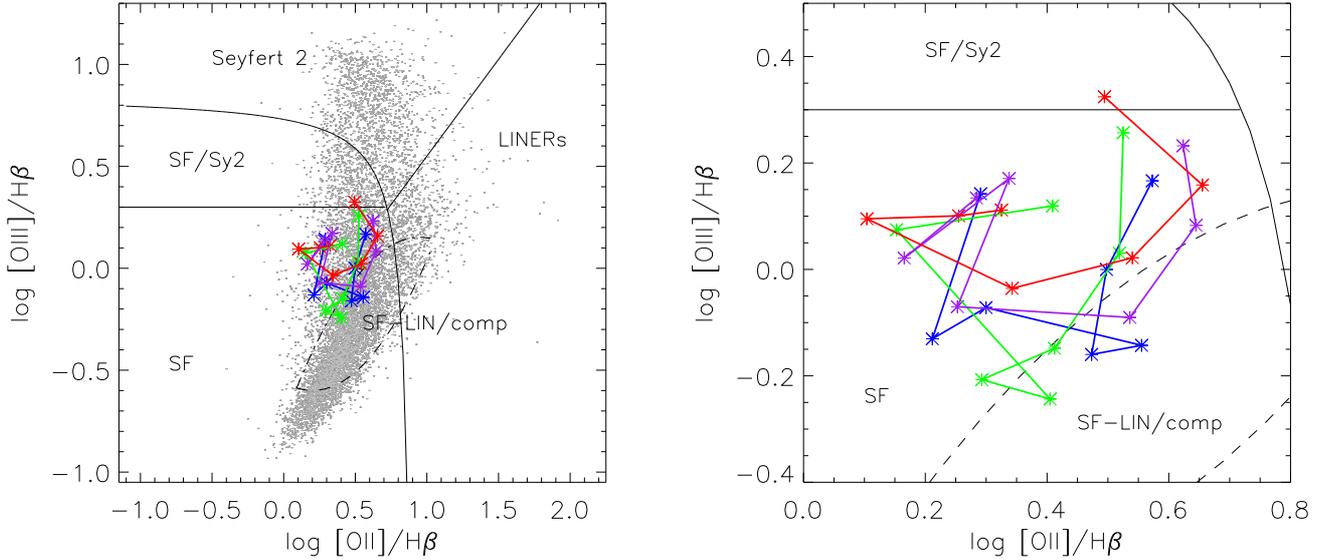}
  \caption{[OII] diagnostic diagram. [OII]/H$\beta$ fux dust-corrected ratios are used. The solid demarcation lines are by \citet{Lamareille2010}. Each color represents the trend with stellar mass for a fixed redshift bin: $0.548<z<0.632$ (blue), $0.632<z<0.716$ (green), $0.716<z<0.8$ (purple) and $0.8<z<0.884$ (red). The total stellar mass increases from left to right. In the left panel, the SDSS galaxies from \citet{Vitale2012} are plotted in gray. The right panel shows a closer view of the zCOSMOS galaxies.}
 \label{oii}
 \end{figure*}

\subsection{High redshift diagnostic diagram ($0.548<z<0.884$)}
For the objects placed at higher redshifts, where we lack useful AGN-activity indicators such as the [NII], [SII] or [OI] lines, a different kind of diagnostic diagram is needed to determine the galaxy spectral type. The \citet{Lamareille2004} involves the [OII] line (Fig. \ref{oii}) and represents a high-redshift option to the more classic red diagnostic diagrams. Unfortunately, this diagram is strongly biased against composites \citep{Lamareille2010}, which overlap with star-forming galaxies and Low Ionization Nuclear Emission Regions \citep[LINERs,][]{Heckman1980}. The [OII] emission line is thought to be either an indicator of ongoing star formation or AGN-activity, due to the relatively low ionization potential that is required compared to the [NII] or [SII] transitions. As a consequence, this diagnostic diagram cannot be considered as reliable as the set of low-redshift diagnostic diagrams \citep{Bongiorno2010,Juneau2011}.\newline
In this diagram it becomes necessary to correct line ratios for reddening. This is because the emission-lines that are used as indicators are placed far away from each other in the spectra and they are thus differently affected by reddening. Another way to overcome this problem is to use the EW of the emission lines. The use of the EW instead than the extinction corrected fluxes does not change our galaxy classification significantly (see Appendix B).\newline
For separating AGN from star-forming galaxies (Fig. \ref{oii}), we used lines which have been empirically defined by \citet{Lamareille2010} by using galaxies from the SDSS. The left panel of Fig. \ref{oii} shows the distribution of SDSS galaxies (in gray) and the zCOSMOS galaxies superimposed on it. For every redshift bin (indicated with different colors) a higher mass means, in most of the cases, a shift towards higher values of [OII]/H$\beta$. Although the stacked galaxies show, as in the [NII] diagram (Fig. \ref{nii}), a trend with the total stellar mass, almost all the objects are placed in the SF zone. For the higher masses, we notice that some of the stacked galaxies fall in the region of mixed contribution from star-forming galaxies and LINERs, while only the points representing the highest mass-redshift bin falls in the star-forming/Seyfert region. This poor ability to properly classify AGN using the [OII]/H$\beta$ ratios was also noticed in the analysis of zCOSMOS spectra for $24$ micron-selected galaxies \citep{Caputi2008,Caputi2009}. The diagram is known to be the most sensitive to shocks. This might be the reason why some objects are likely to be misclassified as star-forming galaxies rather that LINERs, which would appear in a similar position along the y axis, but at higher [OII]/H$\beta$ values.
 \begin{figure} 
  \includegraphics [width=9cm,angle=0]{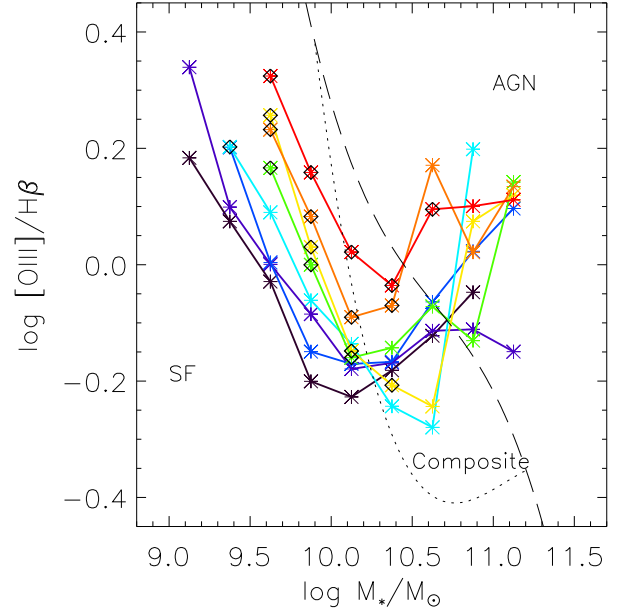}
  \caption{The new diagnostic diagram (MEx) by Juneau et al. (2011) makes use of the stellar mass instead of the [NII]/H$\alpha$ emission line ratio. In this plot, every color represents the evolution in mass for a fixed redshift bin (all the z range): $0.177<z<0.242$ (black), $0.242<z<0.306$ (purple), $0.306<z<0.371$ (blue), $0.371<z<0.436$ (light blue), $0.548<z<0.632$ (green), $0.632<z<0.716$ (yellow), $0.716<z<0.8$ (orange) and $0.8<z<0.884$ (red). The mass increases from left to right along each track. The black diamonds indicate mass incompleteness (see Fig. \ref{binning}).}
 \label{mex}
 \end{figure} 

\subsection{New generation of diagnostic diagrams}
Classical diagnostic diagrams are largely used by the scientific community to classify AGN from emission line ratios, especially in the optical range. Nevertheless, these diagrams present some problems that make it preferable to look for new and more effective diagnostic methods. Recently, \citet{Juneau2011} adopted a new diagram (MEx diagram) that can be used at higher redshift with respect to the classical diagnostic diagrams. A correlation between galaxy stellar mass and metal content has been observed \citep[mass-metallicity relation,][]{Tremonti2004,Savaglio2005} and the [NII]/H$\alpha$ ratio is known to trace metallicity for SFGs \citep{Kewley2006,Stasinska2006}. For this reasons, the authors chose to substitute the emission line ratio [NII]/H$\alpha$ in the BPT diagram with the total stellar mass. This new kind of diagnostics has the important advantage of allowing to classify all the objects of a quite wide redshift range within a unique diagram. Moreover, it avoids some of the issues linked to the reliability of the [OII] diagram, though the MEx diagram suffers from more blending between LINERs and Seyfert 2s with respect to the latter. \newline
The mass trend we already observed in the [NII] diagram is now clear in the MEx diagram (Fig. \ref{mex}), where we can represent galaxies spanning our entire redshift range. As in Fig. \ref{nii} and Fig.\ref{oii}, each color represents a different redshift bin and the points indicate the mass bins. At $\log M_*/M_{\odot}>10.2$, the galaxies leave the SF region to enter the composite region of the diagram. With the increase of the mass, the [OIII]/H$\beta$ ratio decreases up to the point where the galaxies start to be classified as composites. Then, the ratio increases again, as already noted in the [NII] diagram. While the stellar mass tracks overlap in Fig. \ref{nii} and Fig.\ref{oii}, they are parallel to each other in the SF region of the MEx diagram and systematically offset towards the AGN region for increasing redshift (Fig. \ref{mex}). Therefore, higher the redshift higher the number of composites which are classified as transitional objects or AGN. In this respect, the MEx diagram is the one that shows the most clear sign of redshift evolution.

\section{Discussion}
In this section, we discuss our main results and explore future possibilities.
\subsection{Average and median stacking}
Stacking analyses may be challenging to interpret if the stacked galaxies span a broad range of physical properties. When taking averages, galaxies with strong lines might dominate the signal over objects with intrinsically weaker lines. As an example, we would expect massive galaxies that are star-forming and LINERs to have much weaker lines (in particular [OIII]) compared to Seyferts with the same stellar mass. Therefore, a potential risk would be that a few percent of Seyferts dominate the signal and raise [OIII] on the stacked spectrum, yielding a biased view that the stacked galaxies are AGN dominated by number. In order to search for this effect, we compared the results with median stacking and average stacking. We made the comparison for a sub-sample of two median stacked redshift bins ($15$ stacks in total). The line ratios we calculated from the stacks obtained with median and average stacking are very similar and lead to the same spectroscopic classification. Therefore, we did not find any indication of a bias towards more powerful AGN dominating the signal of the stacks. We conclude that there is no substantial difference between median and average stacking of galaxies in our mass-redshift bins. All the measurements shown in this paper are from average stacked spectra. We can identify two possible reasons for the similarity between the measurements: a) From the definition of average and median stacking, we expect the difference between the two to be noticeable only when the number of AGN in a particular mass-z bin is very low (compared to the number of star-forming galaxies). From the results which are shown in the paper and as already expected from previous studies, this is more likely to happen at low masses. However, the chance of noticing a difference between median and average stacking is reduced in case of symmetric shape of the underlying distribution (e.g. Gaussian), when in a large sample the median value of the distribution tends to the average value. In our case, we deal with hundreds of galaxies per bin, thus with a statistics which might be large enough to hide this effect.
b) The effect is still there, but it is not possible to notice the difference between median and average stacking and the trend is somehow lost in the measurements errors, which we estimate to be $<20\%$ (see Section 3.4). The typical difference we find on the emission-line ratios is, in the [NII] diagram, $<5\%$ and $\sim20\%$ for the [NII]/H$\alpha$ and [OIII]/H$\beta$ ratios, respectively. In the [OII] diagram, the difference is $\sim10\%$ and $\sim40\%$ for the [OII]/H$\beta$ and [OIII]/H$\beta$ ratios, respectively.\newline
However, we should keep in mind that in our study all the low redshift mass-redshift bins and several of the high redshift bins are considered to be complete (see Sect. 2.1, Sect. 3.1). Hence, each stack is representative of the average behavior of a particular bin.
\subsection{Visual extinction}
 \begin{figure} 
  \includegraphics [width=9cm,angle=0]{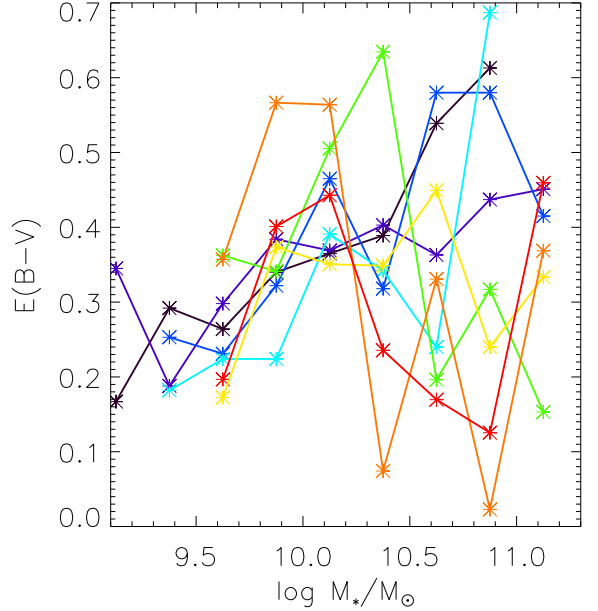}
  \caption{Measured E(B-V) (inferred from the Balmer ratios) as a function of stellar mass, with tracks corresponding to a given redshift: $0.177<z<0.242$ (black), $0.242<z<0.306$ (purple), $0.306<z<0.371$ (blue), $0.371<z<0.436$ (light blue), $0.548<z<0.632$ (green), $0.632<z<0.716$ (yellow), $0.716<z<0.8$ (orange) and $0.8<z<0.884$ (red) (see Fig. \ref{binning}).}
 \label{ebv}
 \end{figure} 
  \begin{figure*} 
   \includegraphics[width=18cm,angle=0]{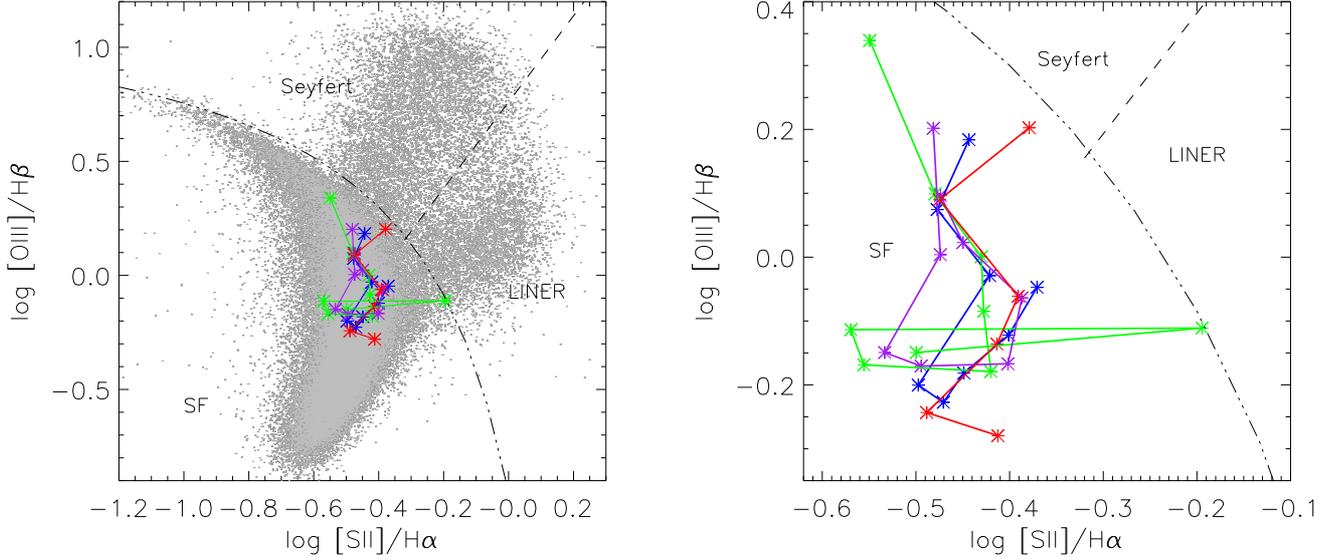}
   \caption{Classical [SII] diagnostic diagram. In both panels demarcation curves by \citet{Kewley2006} separate star-forming galaxies from Seyferts and LINERs. Each color represents the trend with stellar mass for a fixed redshift bin: $0.177<z<0.242$ (blue), $0.242<z<0.306$ (green), $0.306<z<0.371$ (purple) and $0.371<z<0.436$ (red). The total stellar mass increases along each track, starting from the top. In the left panel, the SDSS galaxies from \citet{Vitale2012} are plotted in gray. The right panel offers a closer view of the zCOSMOS galaxies.}
  \label{sii}
  \end{figure*} 
It is known that dust extinction in star-forming galaxies depends upon star formation rate (SFR), metallicity and stellar mass independently \citep{Garn2010}. High extinction values are associated with an increase of any of these three parameters, although the fundamental property connected with extinction seems to be the stellar mass. However, in our study we consider galaxies of different spectral type. We would expect to find increasing values of the visual extinction for higher masses, since the high mass tail of the galaxy mass function of galaxies placed at high redshift ($z\sim 2$) is populated by Ultra Luminous Infrared Galaxies \citep[ULIRGs, infrared luminosity at $8-1000$ microns higher than $10^{12}~L_{\odot}$,][]{Daddi2005} that store a large amount of dust reddening the outcoming light. On the other hand, we may expect the opposite trend, namely a decreasing value of the visual extinction with the stellar mass, since on the high tail of the galaxy mass distribution we observe objects that might have experienced the AGN-feedback (e.g. jets blowing up the gas and dust content). In general, massive systems that have already experienced star formation in the past are found not to contain a large amount of gas and dust. Less massive galaxies, according to the downsizing scenario, are more likely in the evolutionary stage where they still have material to fuel the star-formation.\newline
We find that the distribution of the E(B-V) values is nearly flat for increasing mass and redshift. In particular, the visual extinction measured from the stellar-continuum subtracted spectra shows an increment with the stellar mass only in the low-redshift bins (Fig. \ref{ebv} and Tab. \ref{stellar_pop}). At higher redshifts the same trend may be hidden by the bigger errors in the measurement of E(B-V) (see Sect. 3.4). On the other hand, the E(B-V) values inferred from the SED fitting results displayed in Tab. \ref{stellar_pop} show a different tentative trend. The E(B-V) values are higher in the high-redshift bins, but with no distinction between high and low masses. These evidence are not sufficient to indicate what the main contribution to extinction could be - reddening from dusty ULIRGs or low extinction due to AGN-feedback. The results seems to weakly suggest a higher extinction for the more massive galaxies, especially if placed at higher redshift. However, further studies are needed to shed light on this topic.

\subsection{Diagnostic diagrams}
Emission lines are a good indicator of star-forming or AGN activity. By means of diagnostic diagrams and spectral stacking, we found the indication for a link between the total stellar mass and the chance of identifying galaxies as AGN. A temporal evolution - namely, an observed trend with the redshift - is also possible but, even if present, it is not well visible within the classical diagnostic diagrams. This could be due to our redshift bins being too close to each other for the stacked galaxies to show a trend with redshift in the classification provided by the diagrams. It is important to notice that we are studying a cosmic period ($z<1$) that is considered as crucial for AGN-feedback and quenching of the star formation to happen, as it has seen both the drop in the star formation history and the build up of about half of the red sequence \citep{Bell2007,Faber2007}. However, the time interval over which to probe evolutionary effects is larger when using the MEx diagram ($0.17\lesssim z\lesssim 0.88$) than when using the BPT ($0.17\lesssim z \lesssim 0.44$) or [OII] ($0.55\lesssim z\lesssim 0.88$) diagrams alone.\newline
The indication of an evolutionary trend is present in the MEx diagram (Fig. \ref{mex}). Here the data points representing galaxies in the same mass range are progressively offset to higher values of [OIII]/H$\beta$ for increasing redshift. On the other hand, along each track (fixed redshift, increasing stellar mass) the data show a metal enhancement (see also Fig. \ref{nii}). The reason for the latter could be researched in the mass-metallicity relation and the efficiency of galactic winds in removing metals from low-mass galaxies \citep{Lequeux1979,Tremonti2004,Maiolino2008}. Outflows generated by starburst winds may easily eject metal-enriched gas from low-mass galaxies placed at low redshifts, making their enrichment less significant than in massive systems \citep{Tremonti2004}. Moreover, low mass systems are, according to the “galaxy downsizing” scenario \citep{Juneau2005,Asari2007,Perez-Gonzalez2008} at an earlier evolutionary stage, where they are still converting most of their gas into stars. For this reason they are poorly metal-enriched compared to massive galaxies. However, it should be kept in mind that our purely magnitude-limited sample suffers from selection biases and mass incompleteness towards the highest redshifts, where most of the AGN are found among the stacked galaxies.
The set of classical diagnostic diagrams turned out to be not completely suitable for our study. While the [NII] diagram nicely shows that the mass can determine the galaxy classification, the [SII] - due to the problems already discussed in Sect. 4.1 - does not show the same trend and classifies all the stacked galaxies as SF (Fig. \ref{sii}). The [OII]-based diagram, on the other hand, leaves us with big uncertainties because of the intrinsic nature of the diagram itself (rather sensitive to shocks, it is not able to disentangle the contribution of stellar sources from AGN). As a matter of fact, the diagram shows an overlap between star-forming galaxies and Seyfert 2s, as well as an overlap between the former and composites. However, the higher-mass galaxies are closer to the demarcation line and the AGN region. The usage of the MEx diagnostic diagram partially solves the issues related to the ineffectiveness of the classical diagnostic diagram at high redshift ([OII]) and serves to the purpose of looking for AGN contamination at high-masses (see next section). The mass threshold above which the objects start to be classified as Composite and AGN is well visible.

\subsection{Quenching of the star-formation}
We noticed a general increase of the age of the stellar populations for increasing mass, at both high and low redshifts. Tab. \ref{stellar_pop} summarizes all the information on the stacked galaxy spectra in the stellar mass and redshift bins illustrated in Fig. \ref{binning}. The most suitable stellar continuum template to be subtracted to each stacked spectrum has been chosen from the set of $234$ stellar templates ($39$ BC templates $\times~5$ reddened spectra). The table shows the reddening inferred from the stellar template, E(B-V)$_s$, and the reddening calculated from the Balmer lines, E(B-V)$_m$, whose fluxes have been measured from the stellar-continuum subtracted spectra. The metallicity of the stellar template gives an important indication of the presence of the age-metallicity degeneracy \citep{Worthey1994}. For a fixed redshift and along the mass sequence (Fig. \ref{binning}, Tab. \ref{stellar_pop}), the stellar populations become progressively older. A young stellar population following an older one in the mass sequence presents, in most of the cases, a higher metallicity and a higher reddening as inferred from the stellar fit. A stellar population following one of the same age along the mass sequence is often characterized by equal or higher values of $Z$ and E(B-V)$_s$.\newline
This evidence finds a possible explanation in the mass-assembly downsizing scenario, where most of the massive galaxies build their mass earlier than lower mass galaxies \citep{Cowie1996,Kodama2004,Thomas2005,Tanaka2005,Juneau2005,Bundy2006}. According to this scenario, the number of red massive objects is constant up to $z=1$, dominating as the main contribution to the mass population, while the number density of red galaxies with $M<10^{11}~M_{\odot}$ increases with cosmic time. Conversely, the number density of blue galaxies decreases from $z=1$ to today. Downsizing suggests that something turns the galaxy transformation out at around $z=1$. If the most massive objects build their mass during an earlier period of time with respect to the less massive galaxies, we expect to find older stellar populations in the former because they had more time to evolve.\newline
If the effect of having younger stellar populations at higher redshifts was already expected from different galaxy evolutionary scenarios, the evidence of older stellar populations with increasing mass and fixed redshift points to the downsizing scenario in particular. It is likely that, in these objects, a mechanism acted to quench the star formation. Now the question is: can this process be an AGN-feedback? Considering the downsizing scenario where young low-mass galaxies are still undergoing star formation and high mass galaxies are the product of a earlier mass assembly, the latter are passively-evolving galaxies with no star-formation episodes. In this respect, AGN might have just suppressed star formation with their feedback, by means of transforming blue galaxies into red ones that continue to accrete mass. As a support to this theory, it has been found that the stellar population of the host galaxies appears older after the AGN active phase \citep{Sarzi2005}. Nowadays, several theories and models include the AGN-feedback, but observational evidence are still needed to prove them right.
\begin{landscape} 
 \begin{table}\tiny 
  \caption{Information on the stacked galaxies of the mass-redshift bin of Fig. \ref{binning}. \label{stellar_pop}}
  \centering
    \begin{tabular}{|lr|lr|lr|lr||lr|lr|lr|lr|lr|lr|lr|lr||lr|lr|lr|lr|}
\cline{3-16}
\multicolumn{2}{c|}{ }&SSP 2.5 Gyr &0.2&SSP 1.4 Gyr &0.2&SSP 1.4 Gyr &0.2&SSP 2.5 Gyr&0.2 &SSP 2.5 Gyr &0.2&SSP 1.4 Gyr &0&SSP 1.4 Gyr&0.2\\
\multicolumn{2}{c}{ }&\multicolumn{2}{|c|}{\bf{\textcolor{red}{AGN}}}&\multicolumn{2}{|c|}{\bf{\textcolor{red}{AGN}}}&\multicolumn{2}{|c||}{\bf{\textcolor{red}{AGN}}}&\multicolumn{2}{|c|}{\bf{\textcolor{red}{AGN}}}&\multicolumn{2}{|c|}{\bf{\textcolor{red}{AGN}}}&\multicolumn{2}{c}{\bf{\textcolor{red}{AGN}}}&\multicolumn{2}{|c|}{\bf{\textcolor{red}{AGN}}}\\
\multicolumn{2}{c|}{ }&0.451&0.08&0.415&0.5&0.011&0.5&0.153&0.08&0.333&0.08&0.369&0.5&0.459&0.2\\
\hline
SSP 1.4 Gyr&0.2 &SSP 2.5 Gyr &0.2&SSP 2.5 Gyr &0.2&SSP 11 Gyr &0.2&SSP 1.4 Gyr &0.2&SSP 1.4 Gyr &0.2&SSP 1.4 Gyr &0.2 &SSP 1.4 Gyr &0.2\\
\multicolumn{2}{|c|}{\bf{\textcolor{red}{AGN}}}&\multicolumn{2}{|c|}{\bf{\textcolor{red}{AGN}}}&\multicolumn{2}{|c|}{\bf{\textcolor{red}{AGN}}}&\multicolumn{2}{|c||}{\bf{\textcolor{green}{Comp}}}&\multicolumn{2}{|c|}{\bf{\textcolor{red}{AGN}}}&\multicolumn{2}{|c|}{\bf{\textcolor{red}{AGN}}}&\multicolumn{2}{|c|}{\bf{\textcolor{red}{AGN}}}&\multicolumn{2}{|c|}{\bf{\textcolor{red}{AGN}}}\\
0.613&0.5&0.437&0.2&0.580&0.08&0.687&0.2&0.317&0.2&0.240&0.2&0.023&0.08&0.125&0.08\\
\hline
SSP 2.5 Gyr&0.2 &SSP 1.4 Gyr&0.2&SSP 2.5 Gyr &0.2&SSP 900 Myr &0.2&SSP 1.4 Gyr &0.2&SSP 1.4 Gyr &0.2&Exp 12 Gyr &0.2&SSP 900 Myr &0\\
\multicolumn{2}{|c|}{\bf{\textcolor{green}{Comp}}}&\multicolumn{2}{|c|}{\bf{\textcolor{green}{Comp}}}&\multicolumn{2}{|c|}{\bf{\textcolor{green}{Comp}}}&\multicolumn{2}{|c||}{\bf{\textcolor{green}{Comp}}}&\multicolumn{2}{|c|}{\bf{\textcolor{green}{Comp}}}&\multicolumn{2}{|c|}{\bf{\textcolor{green}{Comp}}}&\multicolumn{2}{|c|}{\bf{\textcolor{red}{AGN}}}&\multicolumn{2}{|c|}{\bf{\textcolor{red}{AGN}}}\\
0.539&0.08&0.363&0.2&0.580&0.08&0.240&0.5&0.196&0.08&0.449&0.08&0.331&0.08&0.170&0.5\\
\hline
Exp 12 Gyr&0.2 &SSP 1.4 Gyr &0.2&SSP 900 Myr &0.2&SSP 900 Myr &0.2&CST 6 Gyr&0.4&CST 6 Gyr &0.4 &Exp 12 Gyr &0.2&Exp 12 Gyr &0.2\\
\multicolumn{2}{|c|}{\bf{\textcolor{blue}{SF}}}&\multicolumn{2}{|c|}{\bf{\textcolor{green}{Comp}}}&\multicolumn{2}{|c|}{\bf{\textcolor{green}{Comp}}}&\multicolumn{2}{|c||}{\bf{\textcolor{blue}{SF}}}&\multicolumn{2}{|c|}{\bf{\textcolor{green}{Comp}}}&\multicolumn{2}{|c|}{\bf{\textcolor{green}{Comp}}}&\multicolumn{2}{|c|}{\bf{\textcolor{green}{Comp}}}&\multicolumn{2}{|c|}{\bf{\textcolor{green}{Comp}}}\\
0.389&0.2&0.403&0.2&0.318&0.5&0.342&0.5&0.635&0.08&0.349&0.08&0.074&0.08&0.235&0.08\\
\hline
SSP 900 Myr &0.2 &SSP 900 Myr &0.2&SSP 1.4 Gyr &0.2&SSP 1.4 Gyr &0.2&CST 6 Gyr &0.4&CST 6 Gyr &0.4 &SSP 25 Myr &0.6&CST 6 Gyr &0.2\\
\multicolumn{2}{|c|}{\bf{\textcolor{blue}{SF}}}&\multicolumn{2}{|c|}{\bf{\textcolor{blue}{SF}}}&\multicolumn{2}{|c|}{\bf{\textcolor{blue}{SF}}}&\multicolumn{2}{|c||}{\bf{\textcolor{blue}{SF}}}&\multicolumn{2}{|c|}{\bf{\textcolor{green}{Comp}}}&\multicolumn{2}{|c|}{\bf{\textcolor{green}{Comp}}}&\multicolumn{2}{|c|}{\bf{\textcolor{green}{Comp}}}&\multicolumn{2}{|c|}{\bf{\textcolor{green}{Comp}}}\\
0.365&0.5&0.369&0.5&0.465&0.08&0.391&0.08&0.505&0.08&0.350&0.08&0.563&0.08&0.443&0.2\\
\hline
Exp 12 Gyr &0.2 &Exp 12 Gyr &0.2&SSP 640 Myr &0.2&SSP 1.4 Gyr &0&CST 6 Gyr &0.2&CST 6 Gyr &0.2 &CST 6 Gyr &0.2&SSP 25 Myr&0.4\\
\multicolumn{2}{|c|}{\bf{\textcolor{blue}{SF}}}&\multicolumn{2}{|c|}{\bf{\textcolor{blue}{SF}}}&\multicolumn{2}{|c|}{\bf{\textcolor{blue}{SF}}}&\multicolumn{2}{|c||}{\bf{\textcolor{blue}{SF}}}&\multicolumn{2}{|c|}{\bf{\textcolor{blue}{SF}}}&\multicolumn{2}{|c|}{\bf{\textcolor{blue}{SF}}}&\multicolumn{2}{|c|}{\bf{\textcolor{blue}{SF}}}&\multicolumn{2}{|c|}{\bf{\textcolor{green}{Comp}}}\\
0.340&0.2&0.384&0.2&0.322&0.5&0.244&0.2&0.341&0.08&0.373&0.08&0.567&0.2&0.401&0.08\\
\hline
SSP 640 Myr&0.2 &SSP 640 Myr &0.2&SSP 900 Myr &0.2&Exp 12 Gyr &0&CST 6 Gyr &0.2&SSP 25 Myr &0.4&SSP 25 Myr &0.4 &CST 6 Gyr &0\\
\multicolumn{2}{|c|}{\bf{\textcolor{blue}{SF}}}&\multicolumn{2}{|c|}{\bf{\textcolor{blue}{SF}}}&\multicolumn{2}{|c|}{\bf{\textcolor{blue}{SF}}}&\multicolumn{2}{|c||}{\bf{\textcolor{blue}{SF}}}&\multicolumn{2}{|c|}{\bf{\textcolor{blue}{SF}}}&\multicolumn{2}{|c|}{\bf{\textcolor{blue}{SF}}}&\multicolumn{2}{|c|}{\bf{\textcolor{blue}{SF}}}&\multicolumn{2}{|c|}{\bf{\textcolor{blue}{SF}}}\\
0.264&0.5&0.298&0.5&0.231&0.08&0.224&0.08&0.362&0.08&0.172&0.08&0.357&0.08&0.197&0.5\\
\hline
SSP 290 Myr &0.4 &SSP 640 Myr &0.2&Exp 12 Gyr&0&SSP 640 Myr &0.2 &\multicolumn{8}{c}{ }\\
\multicolumn{2}{|c|}{\bf{\textcolor{blue}{SF}}}&\multicolumn{2}{|c|}{\bf{\textcolor{blue}{SF}}}&\multicolumn{2}{|c|}{\bf{\textcolor{blue}{SF}}}&\multicolumn{2}{|c||}{\bf{\textcolor{blue}{SF}}} &\multicolumn{8}{c}{ }\\
0.292&0.2&0.188&0.2&0.253&0.08&0.182&0.08&\multicolumn{8}{c}{ }\\
\cline{1-8}\cline{15-16}
SSP 900 Myr&0.2 &SSP 1.4 Gyr &0&\multicolumn{10}{c|}{ }&SP&E(B-V)$_s$\\
\multicolumn{2}{|c|}{\bf{\textcolor{blue}{SF}}}&\multicolumn{2}{|c|}{\bf{\textcolor{blue}{SF}}}&\multicolumn{10}{c}{ }&\multicolumn{2}{|c|}{\bf Class}\\
0.167&0.08&0.345&0.08&\multicolumn{10}{c|}{ }&E(B-V)$_m$&Z\\
\cline{1-4}\cline{15-16}
  \end{tabular}
  \tablefoot{In each box: template of stellar population that fits the stacked zCOSMOS spectrum the best (SP, top-left); reddening of the stellar spectrum (E(B-V)$_s$ top-right); value of the extinction as measured from the Balmer ratios after stellar continuum subtraction (E(B-V)$_m$, bottom-left); metallicity (Z, bottom-right); spectral classification (Class, middle of the box) according to the MEx diagnostic diagram. SSP, CSP and Exp stand for Simple Stellar Population, Constant Stellar Template and Exponential decay of the star-formation, respectively. The legend is to the bottom-right of the table.}
 \end{table}
\end{landscape}

\subsection{AGN identification}
The downsizing scenario takes into account a quenching mechanism to explain galaxy evolution. Our findings fit into the big frame of this latter scenario. Tab. \ref{stellar_pop} reports the spectral classification (AGN, Composite or SF) of the stacked zCOSMOS galaxies according to the MEx diagnostic diagram (Fig. \ref{mex}). It is clearly visible that AGNs are common amongst the high-mass bins. This is particularly the case of higher redshift bins. However, some stacked spectra are placed right on the demarcation curves that separate AGN from Composites, or right beside them. In these cases, it might be either that the stacking galaxies do not belong to a single population - namely, they are a mixture of AGN and star-forming galaxies - or that the underlying population is truly dominated by composite galaxies.\newline
The use of stacked spectra instead of individual spectra leads to some difficulties in estimating the number of SFGs and AGN. If, from one hand, this technique allows to study the average properties of galaxies in a given mass-redshift bin - providing average spectra for otherwise individually unclassifiable galaxies with too low S/N - from the other hand it does not provide us with the exact number of spectral types. Furthermore, each bin collects a variable number of galaxies, due to the presence of filaments (overdensities), voids (underdensities) and selection effects acting at the highest as well as at the lowest stellar masses (Fig. \ref{binning} and Tab. \ref{statistics}). Therefore, it is not possible to compare the exact number of AGN between the different bins. Assuming that all the galaxies in a given stack have the same classification as the stacked spectrum, we obtain the following statistics. The number of single galaxies that are spectroscopically identified as AGN in the high-redshift bins ($5\ 568$ galaxies in total) is $1\ 690$ ($30.3\%$), while the Composites are $2\ 491$ ($44.7\%$). This statistics is higher compared to the galaxies placed in bins at lower redshift ($3\ 742$ galaxies), where we find $395$ AGN ($10.5\%$) and $775$ Composites ($20.7\%$). However, assuming that all the galaxies in a given composite or AGN bin are indeed AGNs will likely result in an overestimate of the true AGN fraction. Nevertheless, the relative difference between the higher and lower redshift bins remain an interesting hint of a greater AGN fraction at earlier epoch. Based on the study of individual galaxy spectra \citet{Ivezic2002} find, with SDSS data, a starburst over AGN ratio equal to $18$, which is a much higher ratio than what we find.\newline
The search for AGN could benefit from the complementary analysis of individual galaxy spectra or from analyses at other wavelengths. The X-domain can identify AGN because the luminous, compact X-ray emission ($L_{2-10~keV}> 10^{42}$ ergs s$^{-1}$) is an almost certain indicator of the existence of an AGN \citep{Bauer2004,Brandt2005,Comastri2008,Brandt2010}, given the extremely small contribution from star formation to the overall emission at these frequencies. X-ray observations have been found to be very efficient in revealing accreting black holes in galaxies which were not classified as AGN from the analysis of optical data \citep[see][and references therein]{Brusa2010}. This suggests that the use of multi-wavelength data can be beneficial in AGN-host studies. The XMM-Newton wide-field survey in the COSMOS field \citep{Hasinger2007} provides a large sample of point-like X-ray sources ($\sim 1\ 800$) with complete ultraviolet to mid-infrared (including Spitzer data) and radio coverage. The survey - thanks to the high efficiency of X-ray observations in identifying AGN - currently allows further investigations \citep[e.g.][]{Brusa2010} and contributes to answering questions on galaxy-black hole coevolution up to high redshifts.

\subsection{Aperture effects}
Aperture biases are particularly important in emission-line studies. Spectroscopic observations are performed using different aperture sizes, depending on the specific instrument. This issue plays a critical role in disentangling the main contribution to galaxy emission. The so-called ``aperture effect'' tells us that the shape and size of the instrument aperture influences the galaxy classification. We would expect big fibers/slits to select emission from all over the galaxy, hence several contributions to the line emission (AGN from the nucleus, starburst from the outer regions of the host galaxy) are present. On the other hand, a smaller aperture would mainly select light coming from the nucleus and the bulge, missing the disk where most of the star formation takes place. For example, SDSS data are known to be affected by aperture effect due to the large fixed size ($3\arcsec$) of the fiber used for observations which includes on average $>20\%$ of galaxies area at z$>0.04$ \citep{Kewley2006}. Therefore, the stellar contribution to the emission from the host causes a large number of sources to be classified as transitional objects in the diagnostic diagrams. The same argument applies at higher redshift, where the slit width becomes comparable to the typical angular size of galaxies. Furthermore, higher-redshift galaxies present higher star-formation rates, so the effect of host galaxy dilution of the AGN signatures could be even more important and yield more often composite signatures. From this point of view, with the $1\arcsec$ width of the slits used for the zCOSMOS observations and a seeing always better than $1.2\arcsec$ \citep{Lilly2007}, our sample is less sensitive to contamination due to emission coming from the most outer regions of the galaxies with respect to other studies at similar redshift.\newline
However, the conventional belief that smaller apertures select more AGN-like emission has been challenged by \citet{Shields2007}. The authors show that for many composite objects identified in the Palomar spectroscopic survey \citep{Ho1997} the observed line-ratios do not appear more AGN-like going on smaller apertures ($10-20$ pc). Therefore, the aperture effect issue remains controversial and it is worth further studies.

\section{Summary and conclusions}
In this work, we used the zCOSMOS 20k catalogue to create a sample of galaxies with reliable spectroscopic redshift and without broad-line AGN, in order avoid overshining issues (AGN brighter than the surrounding galaxy) and investigate the evolution of galaxy spectral properties up to $z\sim 1$. The sample, containig $\sim 1.5\times 10^4$ objects, has been divided in $61$ bins over the stellar mass-redshift plane (Fig. \ref{binning}). This was done taking into account the completeness in mass and the gap in redshift, this latter due to the impossibility of measuring all the emission lines of the classical diagnostic diagrams. In order to improve the signal to noise (S/N) of the single spectra, we have combined all the spectra in each bin to obtain stacks, allowing accurate flux measurements. After stacking the spectra, we performed the stellar continuum subtraction on the galaxy composites, using templates of stellar population synthesis. This provides new spectra that are almost free from absorption contaminations by stars, and also gives information on the age of the stellar populations and their composition (metallicity). Finally, we applied optical emission-lines diagnostics to search for AGN contamination at the highest stellar masses. \newline
Our main findings are:
\begin{itemize}
 \item The visual extinction of the galaxies inferred from the Balmer ratios does not show a strong trend with the total stellar mass all over the z range. It increases slightly at low redshifts for increasing mass, but there is no indication of a clear trend at higher redshifts.
 \item Galaxy stellar populations are older (in terms of time from the last burst of star formation) for more massive objects. This is in agreement with an evolutionary scenario that accounts for a fast mass assembling and stellar population aging in case of massive objects.
 \item In general, galaxy stacked spectra are more likely to be classified as AGN rather that star-forming galaxies for large values of their total stellar mass. In particular, we found that galaxies with $\log M_*/M_{\odot}>10.2$ start to be classified as composite objects (where both photoionization by stars and nuclear activity contribute to produce the emission lines). They are placed in the AGN region of the diagnostic diagrams for the highest masses considered in our sample.
 \item At fixed redshift, the classification of the stacked spectra displays a trend with increasing stellar mass. The general trend follows the left branch of the SDSS classification down to the bulk of the star-forming galaxies population. At the highest stellar masses, the tracks fall in the composite and then in the AGN region. This is visible in the [NII] and in the MEx diagram, whereas it is not well observed in the [OII] diagram.
 \item The MEx diagram represents an useful tool to investigate the ionizing mechanisms inside galaxies, especially at high redshift. The \citet{Lamareille2010} [OII] diagram is not as effective due to the ambiguity between star-forming, composites and Seyfert 2 galaxies in some regions of the diagram. Moreover, there is the need to be in a specific redshift interval to measure all the involved emission lines. The confusion that this diagram shows can be due to its better sensitivity to shocks than to AGN photoionization.
 \item While there is no clear trend with redshift on the low-redshift BPT diagram, there is a trend on the MEx diagram when combining the low- and high-redshift bins. It is visible that the high-redshift stacks display more composite- and AGN-like spectra. However, the explanation of this trend (higher AGN-detection rate at higher redshift) may include true evolutionary effects as well as selection biases.
 \item The link between stellar population age and galaxy stellar mass, combined with the increasing AGN-detection rate for increasing mass and redshift, is consistent with a scenario where AGN could act to quench the star formation and, then, contribute to the transformation from young blue late-type galaxies to old red early-type galaxies.
\end{itemize}  

\begin{acknowledgements} 
The authours thank the anonymous referee for very useful comments and suggestions that helped to improve the paper. M.Vitale is also grateful to M. Garcia-Marin, M. Valencia-S., M. Bremer and A. Eckart for their advice and support. M. Vitale is member of the International Max-Planck Research School (IMPRS) for Astronomy and Astrophysics at the Universities of Bonn and Cologne, supported by the Max-Planck Society. This research is based on observations undertaken at the European Southern Observatory (ESO) Very Large Telescope (VLT) under the Large Program 175.A-0839.
\end{acknowledgements}

\vspace*{0.5cm}
\bibliographystyle{aa} 
\bibliography{bib} 

\begin{thebibliography}{97}
\expandafter\ifx\csname natexlab\endcsname\relax\def\natexlab#1{#1}\fi

\bibitem[{{Aird} {et~al.}(2012){Aird}, {Coil}, {Moustakas}, {Blanton},
  {Burles}, {Cool}, {Eisenstein}, {Smith}, {Wong}, \& {Zhu}}]{Aird2012}
{Aird}, J., {Coil}, A.~L., {Moustakas}, J., {et~al.} 2012, \apj, 746, 90

\bibitem[{{Asari} {et~al.}(2007){Asari}, {Cid Fernandes}, {Stasi{\'n}ska},
  {Torres-Papaqui}, {Mateus}, {Sodr{\'e}}, {Schoenell}, \& {Gomes}}]{Asari2007}
{Asari}, N.~V., {Cid Fernandes}, R., {Stasi{\'n}ska}, G., {et~al.} 2007,
  \mnras, 381, 263

\bibitem[{{Baldry} {et~al.}(2004){Baldry}, {Glazebrook}, {Brinkmann},
  {Ivezi{\'c}}, {Lupton}, {Nichol}, \& {Szalay}}]{Baldry2004}
{Baldry}, I.~K., {Glazebrook}, K., {Brinkmann}, J., {et~al.} 2004, \apj, 600,
  681

\bibitem[{{Baldwin} {et~al.}(1981){Baldwin}, {Phillips}, \&
  {Terlevich}}]{Baldwin1981}
{Baldwin}, J.~A., {Phillips}, M.~M., \& {Terlevich}, R. 1981, \pasp, 93, 5

\bibitem[{{Balogh} {et~al.}(2004){Balogh}, {Baldry}, {Nichol}, {Miller},
  {Bower}, \& {Glazebrook}}]{Balogh2004}
{Balogh}, M.~L., {Baldry}, I.~K., {Nichol}, R., {et~al.} 2004, \apjl, 615, L101

\bibitem[{{Bauer} {et~al.}(2004){Bauer}, {Alexander}, {Brandt}, {Schneider},
  {Treister}, {Hornschemeier}, \& {Garmire}}]{Bauer2004}
{Bauer}, F.~E., {Alexander}, D.~M., {Brandt}, W.~N., {et~al.} 2004, \aj, 128,
  2048

\bibitem[{{Bell} {et~al.}(2007){Bell}, {Zheng}, {Papovich}, {Borch}, {Wolf}, \&
  {Meisenheimer}}]{Bell2007}
{Bell}, E.~F., {Zheng}, X.~Z., {Papovich}, C., {et~al.} 2007, \apj, 663, 834

\bibitem[{{Bertoldi} {et~al.}(2007){Bertoldi}, {Carilli}, {Aravena},
  {Schinnerer}, {Voss}, {Smolcic}, {Jahnke}, {Scoville}, {Blain}, {Menten},
  {Lutz}, {Brusa}, {Taniguchi}, {Capak}, {Mobasher}, {Lilly}, {Thompson},
  {Aussel}, {Kreysa}, {Hasinger}, {Aguirre}, {Schlaerth}, \&
  {Koekemoer}}]{Bertoldi2007}
{Bertoldi}, F., {Carilli}, C., {Aravena}, M., {et~al.} 2007, \apjs, 172, 132

\bibitem[{{Binney}(2004)}]{Binney2004}
{Binney}, J. 2004, in The Riddle of Cooling Flows in Galaxies and Clusters of
  galaxies, ed. T.~{Reiprich}, J.~{Kempner}, \& N.~{Soker}, 233

\bibitem[{{Bolzonella} {et~al.}(2010){Bolzonella}, {Kova{\v c}}, {Pozzetti},
  {Zucca}, {Cucciati}, {Lilly}, {Peng}, {Iovino}, {Zamorani}, {Vergani},
  {Tasca}, {Lamareille}, {Oesch}, {Caputi}, {Kampczyk}, {Bardelli}, {Maier},
  {Abbas}, {Knobel}, {Scodeggio}, {Carollo}, {Contini}, {Kneib}, {Le
  F{\`e}vre}, {Mainieri}, {Renzini}, {Bongiorno}, {Coppa}, {de la Torre}, {de
  Ravel}, {Franzetti}, {Garilli}, {Le Borgne}, {Le Brun}, {Mignoli},
  {Pell{\'o}}, {Perez-Montero}, {Ricciardelli}, {Silverman}, {Tanaka},
  {Tresse}, {Bottini}, {Cappi}, {Cassata}, {Cimatti}, {Guzzo}, {Koekemoer},
  {Leauthaud}, {Maccagni}, {Marinoni}, {McCracken}, {Memeo}, {Meneux},
  {Porciani}, {Scaramella}, {Aussel}, {Capak}, {Halliday}, {Ilbert},
  {Kartaltepe}, {Salvato}, {Sanders}, {Scarlata}, {Scoville}, {Taniguchi}, \&
  {Thompson}}]{Bolzonella2010}
{Bolzonella}, M., {Kova{\v c}}, K., {Pozzetti}, L., {et~al.} 2010, \aap, 524,
  A76

\bibitem[{{Bongiorno} {et~al.}(2010){Bongiorno}, {Mignoli}, {Zamorani},
  {Lamareille}, {Lanzuisi}, {Miyaji}, {Bolzonella}, {Carollo}, {Contini},
  {Kneib}, {Le F{\`e}vre}, {Lilly}, {Mainieri}, {Renzini}, {Scodeggio},
  {Bardelli}, {Brusa}, {Caputi}, {Civano}, {Coppa}, {Cucciati}, {de la Torre},
  {de Ravel}, {Franzetti}, {Garilli}, {Halliday}, {Hasinger}, {Koekemoer},
  {Iovino}, {Kampczyk}, {Knobel}, {Kova{\v c}}, {Le Borgne}, {Le Brun},
  {Maier}, {Merloni}, {Nair}, {Pello}, {Peng}, {Perez Montero}, {Ricciardelli},
  {Salvato}, {Silverman}, {Tanaka}, {Tasca}, {Tresse}, {Vergani}, {Zucca},
  {Abbas}, {Bottini}, {Cappi}, {Cassata}, {Cimatti}, {Guzzo}, {Leauthaud},
  {Maccagni}, {Marinoni}, {McCracken}, {Memeo}, {Meneux}, {Oesch}, {Porciani},
  {Pozzetti}, \& {Scaramella}}]{Bongiorno2010}
{Bongiorno}, A., {Mignoli}, M., {Zamorani}, G., {et~al.} 2010, \aap, 510, A56

\bibitem[{{Brandt} \& {Alexander}(2010)}]{Brandt2010}
{Brandt}, W.~N. \& {Alexander}, D.~M. 2010, Proceedings of the National Academy
  of Science, 107, 7184

\bibitem[{{Brandt} \& {Hasinger}(2005)}]{Brandt2005}
{Brandt}, W.~N. \& {Hasinger}, G. 2005, \araa, 43, 827

\bibitem[{{Brusa} {et~al.}(2010){Brusa}, {Civano}, {Comastri}, {Miyaji},
  {Salvato}, {Zamorani}, {Cappelluti}, {Fiore}, {Hasinger}, {Mainieri},
  {Merloni}, {Bongiorno}, {Capak}, {Elvis}, {Gilli}, {Hao}, {Jahnke},
  {Koekemoer}, {Ilbert}, {Le Floc'h}, {Lusso}, {Mignoli}, {Schinnerer},
  {Silverman}, {Treister}, {Trump}, {Vignali}, {Zamojski}, {Aldcroft},
  {Aussel}, {Bardelli}, {Bolzonella}, {Cappi}, {Caputi}, {Contini},
  {Finoguenov}, {Fruscione}, {Garilli}, {Impey}, {Iovino}, {Iwasawa},
  {Kampczyk}, {Kartaltepe}, {Kneib}, {Knobel}, {Kovac}, {Lamareille},
  {Leborgne}, {Le Brun}, {Le Fevre}, {Lilly}, {Maier}, {McCracken}, {Pello},
  {Peng}, {Perez-Montero}, {de Ravel}, {Sanders}, {Scodeggio}, {Scoville},
  {Tanaka}, {Taniguchi}, {Tasca}, {de la Torre}, {Tresse}, {Vergani}, \&
  {Zucca}}]{Brusa2010}
{Brusa}, M., {Civano}, F., {Comastri}, A., {et~al.} 2010, \apj, 716, 348

\bibitem[{{Bruzual} \& {Charlot}(2003)}]{BruzualCharlot2003}
{Bruzual}, G. \& {Charlot}, S. 2003, \mnras, 344, 1000

\bibitem[{{Bundy} {et~al.}(2006){Bundy}, {Ellis}, {Conselice}, {Taylor},
  {Cooper}, {Willmer}, {Weiner}, {Coil}, {Noeske}, \& {Eisenhardt}}]{Bundy2006}
{Bundy}, K., {Ellis}, R.~S., {Conselice}, C.~J., {et~al.} 2006, \apj, 651, 120

\bibitem[{{Bundy} {et~al.}(2004){Bundy}, {Fukugita}, {Ellis}, {Kodama}, \&
  {Conselice}}]{Bundy2004}
{Bundy}, K., {Fukugita}, M., {Ellis}, R.~S., {Kodama}, T., \& {Conselice},
  C.~J. 2004, \apjl, 601, L123

\bibitem[{{Calzetti} {et~al.}(1994){Calzetti}, {Kinney}, \&
  {Storchi-Bergmann}}]{Calzetti1994}
{Calzetti}, D., {Kinney}, A.~L., \& {Storchi-Bergmann}, T. 1994, \apj, 429, 582

\bibitem[{{Capak} {et~al.}(2007){Capak}, {Aussel}, {Ajiki}, {McCracken},
  {Mobasher}, {Scoville}, {Shopbell}, {Taniguchi}, {Thompson}, {Tribiano},
  {Sasaki}, {Blain}, {Brusa}, {Carilli}, {Comastri}, {Carollo}, {Cassata},
  {Colbert}, {Ellis}, {Elvis}, {Giavalisco}, {Green}, {Guzzo}, {Hasinger},
  {Ilbert}, {Impey}, {Jahnke}, {Kartaltepe}, {Kneib}, {Koda}, {Koekemoer},
  {Komiyama}, {Leauthaud}, {Le Fevre}, {Lilly}, {Liu}, {Massey}, {Miyazaki},
  {Murayama}, {Nagao}, {Peacock}, {Pickles}, {Porciani}, {Renzini}, {Rhodes},
  {Rich}, {Salvato}, {Sanders}, {Scarlata}, {Schiminovich}, {Schinnerer},
  {Scodeggio}, {Sheth}, {Shioya}, {Tasca}, {Taylor}, {Yan}, \&
  {Zamorani}}]{Capak2007}
{Capak}, P., {Aussel}, H., {Ajiki}, M., {et~al.} 2007, \apjs, 172, 99

\bibitem[{{Caputi} {et~al.}(2009){Caputi}, {Lilly}, {Aussel}, {Le Floc'h},
  {Sanders}, {Maier}, {Frayer}, {Carollo}, {Contini}, {Kneib}, {Le F{\`e}vre},
  {Mainieri}, {Renzini}, {Scodeggio}, {Scoville}, {Zamorani}, {Bardelli},
  {Bolzonella}, {Bongiorno}, {Coppa}, {Cucciati}, {de la Torre}, {de Ravel},
  {Franzetti}, {Garilli}, {Ilbert}, {Iovino}, {Kampczyk}, {Kartaltepe},
  {Knobel}, {Kova{\v c}}, {Lamareille}, {Le Borgne}, {Le Brun}, {Mignoli},
  {Peng}, {P{\'e}rez-Montero}, {Ricciardelli}, {Salvato}, {Silverman},
  {Surace}, {Tanaka}, {Tasca}, {Tresse}, {Vergani}, {Zucca}, {Abbas},
  {Bottini}, {Capak}, {Cappi}, {Cassata}, {Cimatti}, {Elvis}, {Hasinger},
  {Koekemoer}, {Leauthaud}, {Maccagni}, {Marinoni}, {McCracken}, {Memeo},
  {Meneux}, {Oesch}, {Pell{\`o}}, {Porciani}, {Pozzetti}, {Scaramella},
  {Scarlata}, {Schiminovich}, {Taniguchi}, \& {Zamojski}}]{Caputi2009}
{Caputi}, K.~I., {Lilly}, S.~J., {Aussel}, H., {et~al.} 2009, \apj, 707, 1387

\bibitem[{{Caputi} {et~al.}(2008){Caputi}, {Lilly}, {Aussel}, {Sanders},
  {Frayer}, {Le F{\`e}vre}, {Renzini}, {Zamorani}, {Scodeggio}, {Contini},
  {Scoville}, {Carollo}, {Hasinger}, {Iovino}, {Le Brun}, {Le Floc'h}, {Maier},
  {Mainieri}, {Mignoli}, {Salvato}, {Schiminovich}, {Silverman}, {Surace},
  {Tasca}, {Abbas}, {Bardelli}, {Bolzonella}, {Bongiorno}, {Bottini}, {Capak},
  {Cappi}, {Cassata}, {Cimatti}, {Cucciati}, {de la Torre}, {de Ravel},
  {Franzetti}, {Fumana}, {Garilli}, {Halliday}, {Ilbert}, {Kampczyk},
  {Kartaltepe}, {Kneib}, {Knobel}, {Kovac}, {Lamareille}, {Leauthaud}, {Le
  Borgne}, {Maccagni}, {Marinoni}, {McCracken}, {Meneux}, {Oesch}, {Pell{\`o}},
  {P{\'e}rez-Montero}, {Porciani}, {Ricciardelli}, {Scaramella}, {Scarlata},
  {Tresse}, {Vergani}, {Walcher}, {Zamojski}, \& {Zucca}}]{Caputi2008}
{Caputi}, K.~I., {Lilly}, S.~J., {Aussel}, H., {et~al.} 2008, \apj, 680, 939

\bibitem[{{Cattaneo} {et~al.}(2009){Cattaneo}, {Faber}, {Binney}, {Dekel},
  {Kormendy}, {Mushotzky}, {Babul}, {Best}, {Br{\"u}ggen}, {Fabian}, {Frenk},
  {Khalatyan}, {Netzer}, {Mahdavi}, {Silk}, {Steinmetz}, \&
  {Wisotzki}}]{Cattaneo2009}
{Cattaneo}, A., {Faber}, S.~M., {Binney}, J., {et~al.} 2009, \nat, 460, 213

\bibitem[{{Chabrier}(2003)}]{Chabrier2003}
{Chabrier}, G. 2003, \apjl, 586, L133

\bibitem[{{Ciotti} \& {Ostriker}(1997)}]{Ciotti1997}
{Ciotti}, L. \& {Ostriker}, J.~P. 1997, \apjl, 487, L105

\bibitem[{{Comastri} \& {Brusa}(2008)}]{Comastri2008}
{Comastri}, A. \& {Brusa}, M. 2008, Astronomische Nachrichten, 329, 122

\bibitem[{{Conselice} {et~al.}(2007){Conselice}, {Bundy}, {Trujillo}, {Coil},
  {Eisenhardt}, {Ellis}, {Georgakakis}, {Huang}, {Lotz}, {Nandra}, {Newman},
  {Papovich}, {Weiner}, \& {Willmer}}]{Conselice2007}
{Conselice}, C.~J., {Bundy}, K., {Trujillo}, I., {et~al.} 2007, \mnras, 381,
  962

\bibitem[{{Cooper} {et~al.}(2008){Cooper}, {Newman}, {Weiner}, {Yan},
  {Willmer}, {Bundy}, {Coil}, {Conselice}, {Davis}, {Faber}, {Gerke},
  {Guhathakurta}, {Koo}, \& {Noeske}}]{Cooper2008}
{Cooper}, M.~C., {Newman}, J.~A., {Weiner}, B.~J., {et~al.} 2008, \mnras, 383,
  1058

\bibitem[{{Cowie} {et~al.}(1996){Cowie}, {Songaila}, {Hu}, \&
  {Cohen}}]{Cowie1996}
{Cowie}, L.~L., {Songaila}, A., {Hu}, E.~M., \& {Cohen}, J.~G. 1996, \aj, 112,
  839

\bibitem[{{Croton} {et~al.}(2006){Croton}, {Springel}, {White}, {De Lucia},
  {Frenk}, {Gao}, {Jenkins}, {Kauffmann}, {Navarro}, \& {Yoshida}}]{Croton2006}
{Croton}, D.~J., {Springel}, V., {White}, S.~D.~M., {et~al.} 2006, \mnras, 365,
  11

\bibitem[{{Daddi} {et~al.}(2005){Daddi}, {Dickinson}, {Chary}, {Pope},
  {Morrison}, {Alexander}, {Bauer}, {Brandt}, {Giavalisco}, {Ferguson}, {Lee},
  {Lehmer}, {Papovich}, \& {Renzini}}]{Daddi2005}
{Daddi}, E., {Dickinson}, M., {Chary}, R., {et~al.} 2005, \apjl, 631, L13

\bibitem[{{Dopita} {et~al.}(2002){Dopita}, {Kewley}, \&
  {Sutherland}}]{Dopita2002}
{Dopita}, M.~A., {Kewley}, L.~J., \& {Sutherland}, R.~S. 2002, in Revista
  Mexicana de Astronomia y Astrofisica, vol. 27, Vol.~12, Revista Mexicana de
  Astronomia y Astrofisica Conference Series, ed. W.~J. {Henney}, J.~{Franco},
  \& M.~{Martos}, 225--229

\bibitem[{{Faber} {et~al.}(2007){Faber}, {Willmer}, {Wolf}, {Koo}, {Weiner},
  {Newman}, {Im}, {Coil}, {Conroy}, {Cooper}, {Davis}, {Finkbeiner}, {Gerke},
  {Gebhardt}, {Groth}, {Guhathakurta}, {Harker}, {Kaiser}, {Kassin},
  {Kleinheinrich}, {Konidaris}, {Kron}, {Lin}, {Luppino}, {Madgwick},
  {Meisenheimer}, {Noeske}, {Phillips}, {Sarajedini}, {Schiavon}, {Simard},
  {Szalay}, {Vogt}, \& {Yan}}]{Faber2007}
{Faber}, S.~M., {Willmer}, C.~N.~A., {Wolf}, C., {et~al.} 2007, \apj, 665, 265

\bibitem[{{Ferrarese} \& {Merritt}(2000)}]{Ferrarese2000}
{Ferrarese}, L. \& {Merritt}, D. 2000, \apjl, 539, L9

\bibitem[{{Filippenko} \& {Ho}(2003)}]{Filippenko2003}
{Filippenko}, A.~V. \& {Ho}, L.~C. 2003, \apjl, 588, L13

\bibitem[{{Garn} \& {Best}(2010)}]{Garn2010}
{Garn}, T. \& {Best}, P.~N. 2010, \mnras, 409, 421

\bibitem[{{Hasinger} {et~al.}(2007){Hasinger}, {Cappelluti}, {Brunner},
  {Brusa}, {Comastri}, {Elvis}, {Finoguenov}, {Fiore}, {Franceschini}, {Gilli},
  {Griffiths}, {Lehmann}, {Mainieri}, {Matt}, {Matute}, {Miyaji}, {Molendi},
  {Paltani}, {Sanders}, {Scoville}, {Tresse}, {Urry}, {Vettolani}, \&
  {Zamorani}}]{Hasinger2007}
{Hasinger}, G., {Cappelluti}, N., {Brunner}, H., {et~al.} 2007, \apjs, 172, 29

\bibitem[{{Heavens} {et~al.}(2004){Heavens}, {Panter}, {Jimenez}, \&
  {Dunlop}}]{Heavens2004}
{Heavens}, A., {Panter}, B., {Jimenez}, R., \& {Dunlop}, J. 2004, \nat, 428,
  625

\bibitem[{{Heckman}(1980)}]{Heckman1980}
{Heckman}, T.~M. 1980, \aap, 87, 152

\bibitem[{{Ho}(2005)}]{Ho2005}
{Ho}, L.~C. 2005, \apj, 629, 680

\bibitem[{{Ho} {et~al.}(1997{\natexlab{a}}){Ho}, {Filippenko}, \&
  {Sargent}}]{Ho1997b}
{Ho}, L.~C., {Filippenko}, A.~V., \& {Sargent}, W.~L.~W. 1997{\natexlab{a}},
  \apj, 487, 568

\bibitem[{{Ho} {et~al.}(1997{\natexlab{b}}){Ho}, {Filippenko}, \&
  {Sargent}}]{Ho1997}
{Ho}, L.~C., {Filippenko}, A.~V., \& {Sargent}, W.~L.~W. 1997{\natexlab{b}},
  \apj, 487, 568

\bibitem[{{Hogg} {et~al.}(2004){Hogg}, {Blanton}, {Brinchmann}, {Eisenstein},
  {Schlegel}, {Gunn}, {McKay}, {Rix}, {Bahcall}, {Brinkmann}, \&
  {Meiksin}}]{Hogg2004}
{Hogg}, D.~W., {Blanton}, M.~R., {Brinchmann}, J., {et~al.} 2004, \apjl, 601,
  L29

\bibitem[{{Hopkins} {et~al.}(2006){Hopkins}, {Hernquist}, {Cox}, {Di Matteo},
  {Robertson}, \& {Springel}}]{Hopkins2006a}
{Hopkins}, P.~F., {Hernquist}, L., {Cox}, T.~J., {et~al.} 2006, \apjs, 163, 1

\bibitem[{{Ilbert} {et~al.}(2009){Ilbert}, {Capak}, {Salvato}, {Aussel},
  {McCracken}, {Sanders}, {Scoville}, {Kartaltepe}, {Arnouts}, {Le Floc'h},
  {Mobasher}, {Taniguchi}, {Lamareille}, {Leauthaud}, {Sasaki}, {Thompson},
  {Zamojski}, {Zamorani}, {Bardelli}, {Bolzonella}, {Bongiorno}, {Brusa},
  {Caputi}, {Carollo}, {Contini}, {Cook}, {Coppa}, {Cucciati}, {de la Torre},
  {de Ravel}, {Franzetti}, {Garilli}, {Hasinger}, {Iovino}, {Kampczyk},
  {Kneib}, {Knobel}, {Kovac}, {Le Borgne}, {Le Brun}, {F{\`e}vre}, {Lilly},
  {Looper}, {Maier}, {Mainieri}, {Mellier}, {Mignoli}, {Murayama}, {Pell{\`o}},
  {Peng}, {P{\'e}rez-Montero}, {Renzini}, {Ricciardelli}, {Schiminovich},
  {Scodeggio}, {Shioya}, {Silverman}, {Surace}, {Tanaka}, {Tasca}, {Tresse},
  {Vergani}, \& {Zucca}}]{Ilbert2009}
{Ilbert}, O., {Capak}, P., {Salvato}, M., {et~al.} 2009, \apj, 690, 1236

\bibitem[{{Ivezi{\'c}} {et~al.}(2002){Ivezi{\'c}}, {Menou}, {Knapp}, {Strauss},
  {Lupton}, {Vanden Berk}, {Richards}, {Tremonti}, {Weinstein}, {Anderson},
  {Bahcall}, {Becker}, {Bernardi}, {Blanton}, {Eisenstein}, {Fan},
  {Finkbeiner}, {Finlator}, {Frieman}, {Gunn}, {Hall}, {Kim}, {Kinkhabwala},
  {Narayanan}, {Rockosi}, {Schlegel}, {Schneider}, {Strateva}, {SubbaRao},
  {Thakar}, {Voges}, {White}, {Yanny}, {Brinkmann}, {Doi}, {Fukugita},
  {Hennessy}, {Munn}, {Nichol}, \& {York}}]{Ivezic2002}
{Ivezi{\'c}}, {\v Z}., {Menou}, K., {Knapp}, G.~R., {et~al.} 2002, \aj, 124,
  2364

\bibitem[{{Jimenez} {et~al.}(2005){Jimenez}, {Panter}, {Heavens}, \&
  {Verde}}]{Jimenez2005}
{Jimenez}, R., {Panter}, B., {Heavens}, A.~F., \& {Verde}, L. 2005, \mnras,
  356, 495

\bibitem[{{Juneau} {et~al.}(2011){Juneau}, {Dickinson}, {Alexander}, \&
  {Salim}}]{Juneau2011}
{Juneau}, S., {Dickinson}, M., {Alexander}, D.~M., \& {Salim}, S. 2011, \apj,
  736, 104

\bibitem[{{Juneau} {et~al.}(2005){Juneau}, {Glazebrook}, {Crampton},
  {McCarthy}, {Savaglio}, {Abraham}, {Carlberg}, {Chen}, {Le Borgne}, {Marzke},
  {Roth}, {J{\o}rgensen}, {Hook}, \& {Murowinski}}]{Juneau2005}
{Juneau}, S., {Glazebrook}, K., {Crampton}, D., {et~al.} 2005, \apjl, 619, L135

\bibitem[{{Kauffmann} {et~al.}(2003){Kauffmann}, {Heckman}, {Tremonti},
  {Brinchmann}, {Charlot}, {White}, {Ridgway}, {Brinkmann}, {Fukugita}, {Hall},
  {Ivezi{\'c}}, {Richards}, \& {Schneider}}]{Kauffmann2003a}
{Kauffmann}, G., {Heckman}, T.~M., {Tremonti}, C., {et~al.} 2003, \mnras, 346,
  1055

\bibitem[{{Kewley} {et~al.}(2001){Kewley}, {Dopita}, {Sutherland}, {Heisler},
  \& {Trevena}}]{Kewley2001}
{Kewley}, L.~J., {Dopita}, M.~A., {Sutherland}, R.~S., {Heisler}, C.~A., \&
  {Trevena}, J. 2001, \apj, 556, 121

\bibitem[{{Kewley} {et~al.}(2003){Kewley}, {Geller}, \& {Jansen}}]{Kewley2003}
{Kewley}, L.~J., {Geller}, M.~J., \& {Jansen}, R.~A. 2003, in Bulletin of the
  American Astronomical Society, Vol.~35, American Astronomical Society Meeting
  Abstracts, 119.01

\bibitem[{{Kewley} {et~al.}(2006){Kewley}, {Groves}, {Kauffmann}, \&
  {Heckman}}]{Kewley2006}
{Kewley}, L.~J., {Groves}, B., {Kauffmann}, G., \& {Heckman}, T. 2006, \mnras,
  372, 961

\bibitem[{{Kodama} {et~al.}(2004){Kodama}, {Yamada}, {Akiyama}, {Aoki}, {Doi},
  {Furusawa}, {Fuse}, {Imanishi}, {Ishida}, {Iye}, {Kajisawa}, {Karoji},
  {Kobayashi}, {Komiyama}, {Kosugi}, {Maeda}, {Miyazaki}, {Mizumoto},
  {Morokuma}, {Nakata}, {Noumaru}, {Ogasawara}, {Ouchi}, {Sasaki}, {Sekiguchi},
  {Shimasaku}, {Simpson}, {Takata}, {Tanaka}, {Ueda}, {Yasuda}, \&
  {Yoshida}}]{Kodama2004}
{Kodama}, T., {Yamada}, T., {Akiyama}, M., {et~al.} 2004, \mnras, 350, 1005

\bibitem[{{Koekemoer} {et~al.}(2007){Koekemoer}, {Aussel}, {Calzetti}, {Capak},
  {Giavalisco}, {Kneib}, {Leauthaud}, {Le F{\`e}vre}, {McCracken}, {Massey},
  {Mobasher}, {Rhodes}, {Scoville}, \& {Shopbell}}]{Koekemoer2007}
{Koekemoer}, A.~M., {Aussel}, H., {Calzetti}, D., {et~al.} 2007, \apjs, 172,
  196

\bibitem[{{Lamareille}(2010)}]{Lamareille2010}
{Lamareille}, F. 2010, \aap, 509, A53

\bibitem[{{Lamareille} {et~al.}(2004){Lamareille}, {Mouhcine}, {Contini},
  {Lewis}, \& {Maddox}}]{Lamareille2004}
{Lamareille}, F., {Mouhcine}, M., {Contini}, T., {Lewis}, I., \& {Maddox}, S.
  2004, \mnras, 350, 396

\bibitem[{{Lequeux} {et~al.}(1979){Lequeux}, {Peimbert}, {Rayo}, {Serrano}, \&
  {Torres-Peimbert}}]{Lequeux1979}
{Lequeux}, J., {Peimbert}, M., {Rayo}, J.~F., {Serrano}, A., \&
  {Torres-Peimbert}, S. 1979, \aap, 80, 155

\bibitem[{{Lilly} {et~al.}(2009){Lilly}, {Le Brun}, {Maier}, {Mainieri},
  {Mignoli}, {Scodeggio}, {Zamorani}, {Carollo}, {Contini}, {Kneib}, {Le
  F{\`e}vre}, {Renzini}, {Bardelli}, {Bolzonella}, {Bongiorno}, {Caputi},
  {Coppa}, {Cucciati}, {de la Torre}, {de Ravel}, {Franzetti}, {Garilli},
  {Iovino}, {Kampczyk}, {Kovac}, {Knobel}, {Lamareille}, {Le Borgne}, {Pello},
  {Peng}, {P{\'e}rez-Montero}, {Ricciardelli}, {Silverman}, {Tanaka}, {Tasca},
  {Tresse}, {Vergani}, {Zucca}, {Ilbert}, {Salvato}, {Oesch}, {Abbas},
  {Bottini}, {Capak}, {Cappi}, {Cassata}, {Cimatti}, {Elvis}, {Fumana},
  {Guzzo}, {Hasinger}, {Koekemoer}, {Leauthaud}, {Maccagni}, {Marinoni},
  {McCracken}, {Memeo}, {Meneux}, {Porciani}, {Pozzetti}, {Sanders},
  {Scaramella}, {Scarlata}, {Scoville}, {Shopbell}, \& {Taniguchi}}]{Lilly2009}
{Lilly}, S.~J., {Le Brun}, V., {Maier}, C., {et~al.} 2009, \apjs, 184, 218

\bibitem[{{Lilly} {et~al.}(2007){Lilly}, {Le F{\`e}vre}, {Renzini}, {Zamorani},
  {Scodeggio}, {Contini}, {Carollo}, {Hasinger}, {Kneib}, {Iovino}, {Le Brun},
  {Maier}, {Mainieri}, {Mignoli}, {Silverman}, {Tasca}, {Bolzonella},
  {Bongiorno}, {Bottini}, {Capak}, {Caputi}, {Cimatti}, {Cucciati}, {Daddi},
  {Feldmann}, {Franzetti}, {Garilli}, {Guzzo}, {Ilbert}, {Kampczyk}, {Kovac},
  {Lamareille}, {Leauthaud}, {Borgne}, {McCracken}, {Marinoni}, {Pello},
  {Ricciardelli}, {Scarlata}, {Vergani}, {Sanders}, {Schinnerer}, {Scoville},
  {Taniguchi}, {Arnouts}, {Aussel}, {Bardelli}, {Brusa}, {Cappi}, {Ciliegi},
  {Finoguenov}, {Foucaud}, {Franceschini}, {Halliday}, {Impey}, {Knobel},
  {Koekemoer}, {Kurk}, {Maccagni}, {Maddox}, {Marano}, {Marconi}, {Meneux},
  {Mobasher}, {Moreau}, {Peacock}, {Porciani}, {Pozzetti}, {Scaramella},
  {Schiminovich}, {Shopbell}, {Smail}, {Thompson}, {Tresse}, {Vettolani},
  {Zanichelli}, \& {Zucca}}]{Lilly2007}
{Lilly}, S.~J., {Le F{\`e}vre}, O., {Renzini}, A., {et~al.} 2007, \apjs, 172,
  70

\bibitem[{{Magorrian} {et~al.}(1998){Magorrian}, {Tremaine}, {Richstone},
  {Bender}, {Bower}, {Dressler}, {Faber}, {Gebhardt}, {Green}, {Grillmair},
  {Kormendy}, \& {Lauer}}]{Magorrian1998}
{Magorrian}, J., {Tremaine}, S., {Richstone}, D., {et~al.} 1998, \aj, 115, 2285

\bibitem[{{Maiolino} {et~al.}(2008){Maiolino}, {Nagao}, {Grazian}, {Cocchia},
  {Marconi}, {Mannucci}, {Cimatti}, {Pipino}, {Ballero}, {Calura}, {Chiappini},
  {Fontana}, {Granato}, {Matteucci}, {Pastorini}, {Pentericci}, {Risaliti},
  {Salvati}, \& {Silva}}]{Maiolino2008}
{Maiolino}, R., {Nagao}, T., {Grazian}, A., {et~al.} 2008, \aap, 488, 463

\bibitem[{{Martig} {et~al.}(2009){Martig}, {Bournaud}, {Teyssier}, \&
  {Dekel}}]{Martig2009}
{Martig}, M., {Bournaud}, F., {Teyssier}, R., \& {Dekel}, A. 2009, \apj, 707,
  250

\bibitem[{{McCracken} {et~al.}(2010){McCracken}, {Capak}, {Salvato}, {Aussel},
  {Thompson}, {Daddi}, {Sanders}, {Kneib}, {Willott}, {Mancini}, {Renzini},
  {Cook}, {Le F{\`e}vre}, {Ilbert}, {Kartaltepe}, {Koekemoer}, {Mellier},
  {Murayama}, {Scoville}, {Shioya}, \& {Tanaguchi}}]{McCracken2010}
{McCracken}, H.~J., {Capak}, P., {Salvato}, M., {et~al.} 2010, \apj, 708, 202

\bibitem[{{Miller} {et~al.}(2003){Miller}, {Nichol}, {G{\'o}mez}, {Hopkins}, \&
  {Bernardi}}]{Miller2003}
{Miller}, C.~J., {Nichol}, R.~C., {G{\'o}mez}, P.~L., {Hopkins}, A.~M., \&
  {Bernardi}, M. 2003, \apj, 597, 142

\bibitem[{{Narayanan} {et~al.}(2008){Narayanan}, {Cox}, {Shirley}, {Dav{\'e}},
  {Hernquist}, \& {Walker}}]{Narayanan2008}
{Narayanan}, D., {Cox}, T.~J., {Shirley}, Y., {et~al.} 2008, \apj, 684, 996

\bibitem[{{Osterbrock}(1989)}]{Osterbrock1989}
{Osterbrock}, D.~E. 1989, {Astrophysics of gaseous nebulae and active galactic
  nuclei}

\bibitem[{{Padovani} \& {Urry}(1992)}]{Padovani1992}
{Padovani}, P. \& {Urry}, C.~M. 1992, \apj, 387, 449

\bibitem[{{P{\'e}rez-Gonz{\'a}lez} {et~al.}(2008){P{\'e}rez-Gonz{\'a}lez},
  {Rieke}, {Villar}, {Barro}, {Blaylock}, {Egami}, {Gallego}, {Gil de Paz},
  {Pascual}, {Zamorano}, \& {Donley}}]{Perez-Gonzalez2008}
{P{\'e}rez-Gonz{\'a}lez}, P.~G., {Rieke}, G.~H., {Villar}, V., {et~al.} 2008,
  \apj, 675, 234

\bibitem[{{P{\'e}rez-Montero} {et~al.}(2007){P{\'e}rez-Montero}, {H{\"a}gele},
  {Contini}, \& {D{\'{\i}}az}}]{Perez-Montero2007}
{P{\'e}rez-Montero}, E., {H{\"a}gele}, G.~F., {Contini}, T., \& {D{\'{\i}}az},
  {\'A}.~I. 2007, \mnras, 381, 125

\bibitem[{{Pozzetti} {et~al.}(2010){Pozzetti}, {Bolzonella}, {Zucca},
  {Zamorani}, {Lilly}, {Renzini}, {Moresco}, {Mignoli}, {Cassata}, {Tasca},
  {Lamareille}, {Maier}, {Meneux}, {Halliday}, {Oesch}, {Vergani}, {Caputi},
  {Kova{\v c}}, {Cimatti}, {Cucciati}, {Iovino}, {Peng}, {Carollo}, {Contini},
  {Kneib}, {Le F{\'e}vre}, {Mainieri}, {Scodeggio}, {Bardelli}, {Bongiorno},
  {Coppa}, {de la Torre}, {de Ravel}, {Franzetti}, {Garilli}, {Kampczyk},
  {Knobel}, {Le Borgne}, {Le Brun}, {Pell{\`o}}, {Perez Montero},
  {Ricciardelli}, {Silverman}, {Tanaka}, {Tresse}, {Abbas}, {Bottini}, {Cappi},
  {Guzzo}, {Koekemoer}, {Leauthaud}, {Maccagni}, {Marinoni}, {McCracken},
  {Memeo}, {Porciani}, {Scaramella}, {Scarlata}, \& {Scoville}}]{Pozzetti2010}
{Pozzetti}, L., {Bolzonella}, M., {Zucca}, E., {et~al.} 2010, \aap, 523, A13

\bibitem[{{Richstone} {et~al.}(1998){Richstone}, {Ajhar}, {Bender}, {Bower},
  {Dressler}, {Faber}, {Filippenko}, {Gebhardt}, {Green}, {Ho}, {Kormendy},
  {Lauer}, {Magorrian}, \& {Tremaine}}]{Richstone1998}
{Richstone}, D., {Ajhar}, E.~A., {Bender}, R., {et~al.} 1998, \nat, 395, A14

\bibitem[{{Rola} {et~al.}(1997){Rola}, {Terlevich}, \& {Terlevich}}]{Rola1997}
{Rola}, C.~S., {Terlevich}, E., \& {Terlevich}, R.~J. 1997, \mnras, 289, 419

\bibitem[{{Sanders} {et~al.}(2007){Sanders}, {Salvato}, {Aussel}, {Ilbert},
  {Scoville}, {Surace}, {Frayer}, {Sheth}, {Helou}, {Brooke}, {Bhattacharya},
  {Yan}, {Kartaltepe}, {Barnes}, {Blain}, {Calzetti}, {Capak}, {Carilli},
  {Carollo}, {Comastri}, {Daddi}, {Ellis}, {Elvis}, {Fall}, {Franceschini},
  {Giavalisco}, {Hasinger}, {Impey}, {Koekemoer}, {Le F{\`e}vre}, {Lilly},
  {Liu}, {McCracken}, {Mobasher}, {Renzini}, {Rich}, {Schinnerer}, {Shopbell},
  {Taniguchi}, {Thompson}, {Urry}, \& {Williams}}]{Sanders2007}
{Sanders}, D.~B., {Salvato}, M., {Aussel}, H., {et~al.} 2007, \apjs, 172, 86

\bibitem[{{Sanders} {et~al.}(1988){Sanders}, {Soifer}, {Elias}, {Neugebauer},
  \& {Matthews}}]{Sanders1988b}
{Sanders}, D.~B., {Soifer}, B.~T., {Elias}, J.~H., {Neugebauer}, G., \&
  {Matthews}, K. 1988, \apjl, 328, L35

\bibitem[{{Sarzi} {et~al.}(2005){Sarzi}, {Rix}, {Shields}, {Ho}, {Barth},
  {Rudnick}, {Filippenko}, \& {Sargent}}]{Sarzi2005}
{Sarzi}, M., {Rix}, H.-W., {Shields}, J.~C., {et~al.} 2005, \apj, 628, 169

\bibitem[{{Savaglio} {et~al.}(2005){Savaglio}, {Glazebrook}, {Le Borgne},
  {Juneau}, {Abraham}, {Chen}, {Crampton}, {McCarthy}, {Carlberg}, {Marzke},
  {Roth}, {J{\o}rgensen}, \& {Murowinski}}]{Savaglio2005}
{Savaglio}, S., {Glazebrook}, K., {Le Borgne}, D., {et~al.} 2005, \apj, 635,
  260

\bibitem[{{Schawinski} {et~al.}(2007){Schawinski}, {Thomas}, {Sarzi},
  {Maraston}, {Kaviraj}, {Joo}, {Yi}, \& {Silk}}]{Schawinski2007}
{Schawinski}, K., {Thomas}, D., {Sarzi}, M., {et~al.} 2007, \mnras, 382, 1415

\bibitem[{{Schawinski} {et~al.}(2010){Schawinski}, {Urry}, {Virani}, {Coppi},
  {Bamford}, {Treister}, {Lintott}, {Sarzi}, {Keel}, {Kaviraj}, {Cardamone},
  {Masters}, {Ross}, {Andreescu}, {Murray}, {Nichol}, {Raddick}, {Slosar},
  {Szalay}, {Thomas}, \& {Vandenberg}}]{Schawinski2010}
{Schawinski}, K., {Urry}, C.~M., {Virani}, S., {et~al.} 2010, \apj, 711, 284

\bibitem[{{Schinnerer} {et~al.}(2007){Schinnerer}, {Smol{\v c}i{\'c}},
  {Carilli}, {Bondi}, {Ciliegi}, {Jahnke}, {Scoville}, {Aussel}, {Bertoldi},
  {Blain}, {Impey}, {Koekemoer}, {Le Fevre}, \& {Urry}}]{Schinnerer2007}
{Schinnerer}, E., {Smol{\v c}i{\'c}}, V., {Carilli}, C.~L., {et~al.} 2007,
  \apjs, 172, 46

\bibitem[{{Scoville} {et~al.}(2007){Scoville}, {Aussel}, {Brusa}, {Capak},
  {Carollo}, {Elvis}, {Giavalisco}, {Guzzo}, {Hasinger}, {Impey}, {Kneib},
  {LeFevre}, {Lilly}, {Mobasher}, {Renzini}, {Rich}, {Sanders}, {Schinnerer},
  {Schminovich}, {Shopbell}, {Taniguchi}, \& {Tyson}}]{Scoville2007}
{Scoville}, N., {Aussel}, H., {Brusa}, M., {et~al.} 2007, \apjs, 172, 1

\bibitem[{{Shields} {et~al.}(2007){Shields}, {Rix}, {Sarzi}, {Barth},
  {Filippenko}, {Ho}, {McIntosh}, {Rudnick}, \& {Sargent}}]{Shields2007}
{Shields}, J.~C., {Rix}, H.-W., {Sarzi}, M., {et~al.} 2007, \apj, 654, 125

\bibitem[{{Silk}(2005)}]{Silk2005}
{Silk}, J. 2005, \mnras, 364, 1337

\bibitem[{{Silk} \& {Rees}(1998)}]{Silk1998}
{Silk}, J. \& {Rees}, M.~J. 1998, \aap, 331, L1

\bibitem[{{Silverman} {et~al.}(2009){Silverman}, {Lamareille}, {Maier},
  {Lilly}, {Mainieri}, {Brusa}, {Cappelluti}, {Hasinger}, {Zamorani},
  {Scodeggio}, {Bolzonella}, {Contini}, {Carollo}, {Jahnke}, {Kneib}, {Le
  F{\`e}vre}, {Merloni}, {Bardelli}, {Bongiorno}, {Brunner}, {Caputi},
  {Civano}, {Comastri}, {Coppa}, {Cucciati}, {de la Torre}, {de Ravel},
  {Elvis}, {Finoguenov}, {Fiore}, {Franzetti}, {Garilli}, {Gilli}, {Iovino},
  {Kampczyk}, {Knobel}, {Kova{\v c}}, {Le Borgne}, {Le Brun}, {Mignoli},
  {Pello}, {Peng}, {Perez Montero}, {Ricciardelli}, {Tanaka}, {Tasca},
  {Tresse}, {Vergani}, {Vignali}, {Zucca}, {Bottini}, {Cappi}, {Cassata},
  {Fumana}, {Griffiths}, {Kartaltepe}, {Koekemoer}, {Marinoni}, {McCracken},
  {Memeo}, {Meneux}, {Oesch}, {Porciani}, \& {Salvato}}]{Silverman2009}
{Silverman}, J.~D., {Lamareille}, F., {Maier}, C., {et~al.} 2009, \apj, 696,
  396

\bibitem[{{Springel} {et~al.}(2005){Springel}, {Di Matteo}, \&
  {Hernquist}}]{Springel2005a}
{Springel}, V., {Di Matteo}, T., \& {Hernquist}, L. 2005, \apjl, 620, L79

\bibitem[{{Stasi{\'n}ska} {et~al.}(2006){Stasi{\'n}ska}, {Cid Fernandes},
  {Mateus}, {Sodr{\'e}}, \& {Asari}}]{Stasinska2006}
{Stasi{\'n}ska}, G., {Cid Fernandes}, R., {Mateus}, A., {Sodr{\'e}}, L., \&
  {Asari}, N.~V. 2006, \mnras, 371, 972

\bibitem[{{Strateva} {et~al.}(2001){Strateva}, {Ivezi{\'c}}, {Knapp},
  {Narayanan}, {Strauss}, {Gunn}, {Lupton}, {Schlegel}, {Bahcall}, {Brinkmann},
  {Brunner}, {Budav{\'a}ri}, {Csabai}, {Castander}, {Doi}, {Fukugita}, {Gy{\H
  o}ry}, {Hamabe}, {Hennessy}, {Ichikawa}, {Kunszt}, {Lamb}, {McKay},
  {Okamura}, {Racusin}, {Sekiguchi}, {Schneider}, {Shimasaku}, \&
  {York}}]{Strateva2001}
{Strateva}, I., {Ivezi{\'c}}, {\v Z}., {Knapp}, G.~R., {et~al.} 2001, \aj, 122,
  1861

\bibitem[{{Tanaka} {et~al.}(2005){Tanaka}, {Kodama}, {Arimoto}, {Okamura},
  {Umetsu}, {Shimasaku}, {Tanaka}, \& {Yamada}}]{Tanaka2005}
{Tanaka}, M., {Kodama}, T., {Arimoto}, N., {et~al.} 2005, \mnras, 362, 268

\bibitem[{{Taniguchi} {et~al.}(2007){Taniguchi}, {Scoville}, {Murayama},
  {Sanders}, {Mobasher}, {Aussel}, {Capak}, {Ajiki}, {Miyazaki}, {Komiyama},
  {Shioya}, {Nagao}, {Sasaki}, {Koda}, {Carilli}, {Giavalisco}, {Guzzo},
  {Hasinger}, {Impey}, {LeFevre}, {Lilly}, {Renzini}, {Rich}, {Schinnerer},
  {Shopbell}, {Kaifu}, {Karoji}, {Arimoto}, {Okamura}, \&
  {Ohta}}]{Taniguchi2007}
{Taniguchi}, Y., {Scoville}, N., {Murayama}, T., {et~al.} 2007, \apjs, 172, 9

\bibitem[{{Thomas} {et~al.}(2005){Thomas}, {Maraston}, {Bender}, \& {Mendes de
  Oliveira}}]{Thomas2005}
{Thomas}, D., {Maraston}, C., {Bender}, R., \& {Mendes de Oliveira}, C. 2005,
  \apj, 621, 673

\bibitem[{{Tremonti} {et~al.}(2004){Tremonti}, {Heckman}, {Kauffmann},
  {Brinchmann}, {Charlot}, {White}, {Seibert}, {Peng}, {Schlegel}, {Uomoto},
  {Fukugita}, \& {Brinkmann}}]{Tremonti2004}
{Tremonti}, C.~A., {Heckman}, T.~M., {Kauffmann}, G., {et~al.} 2004, \apj, 613,
  898

\bibitem[{{Tresse} {et~al.}(1996){Tresse}, {Rola}, {Hammer}, {Stasi{\'n}ska},
  {Le Fevre}, {Lilly}, \& {Crampton}}]{Tresse1996}
{Tresse}, L., {Rola}, C., {Hammer}, F., {et~al.} 1996, \mnras, 281, 847

\bibitem[{{Treu} {et~al.}(2005){Treu}, {Ellis}, {Liao}, \& {van
  Dokkum}}]{Treu2005}
{Treu}, T., {Ellis}, R.~S., {Liao}, T.~X., \& {van Dokkum}, P.~G. 2005, \apjl,
  622, L5

\bibitem[{{van der Wel} {et~al.}(2005){van der Wel}, {Franx}, {van Dokkum},
  {Rix}, {Illingworth}, \& {Rosati}}]{vanderWel2005}
{van der Wel}, A., {Franx}, M., {van Dokkum}, P.~G., {et~al.} 2005, \apj, 631,
  145

\bibitem[{{Veilleux} \& {Osterbrock}(1987)}]{Veilleux1987}
{Veilleux}, S. \& {Osterbrock}, D.~E. 1987, \apjs, 63, 295

\bibitem[{{Vitale} {et~al.}(2012){Vitale}, {Zuther},
  {Garc{\'{\i}}a-Mar{\'{\i}}n}, {Eckart}, {Bremer}, {Valencia-S.}, \&
  {Zensus}}]{Vitale2012}
{Vitale}, M., {Zuther}, J., {Garc{\'{\i}}a-Mar{\'{\i}}n}, M., {et~al.} 2012,
  \aap, 546, A17

\bibitem[{{Worthey}(1994)}]{Worthey1994}
{Worthey}, G. 1994, \apjs, 95, 107

\end{thebibliography}

\begin{appendix}
\section \newline
Here we present the zCOSMOS stacked spectra which have been used in our study.
\begin{figure*} 
  \centering
  \includegraphics[width=17cm,angle=0]{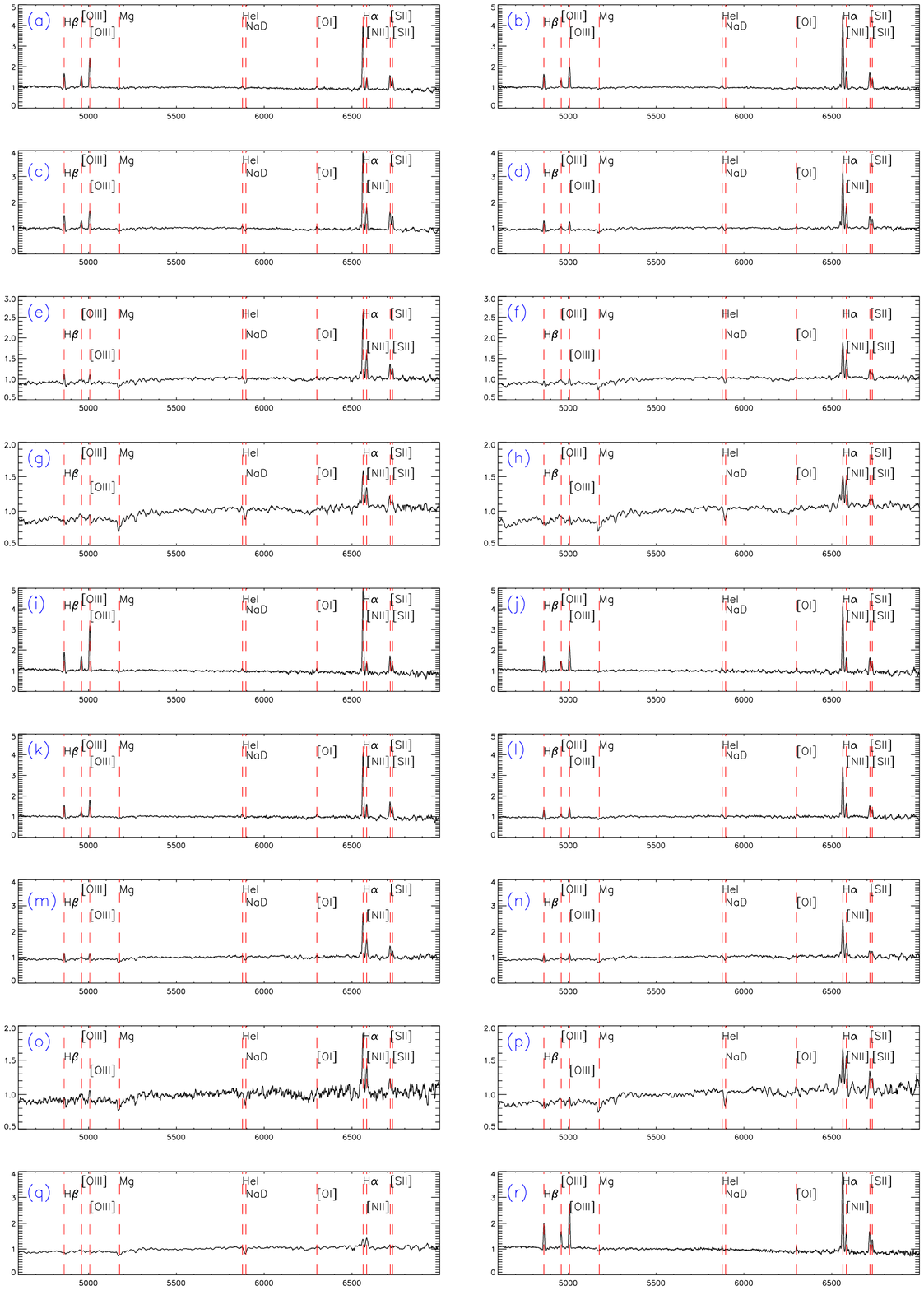}
\end{figure*}

\begin{figure*} 
  \centering
  \includegraphics[width=17cm,angle=0]{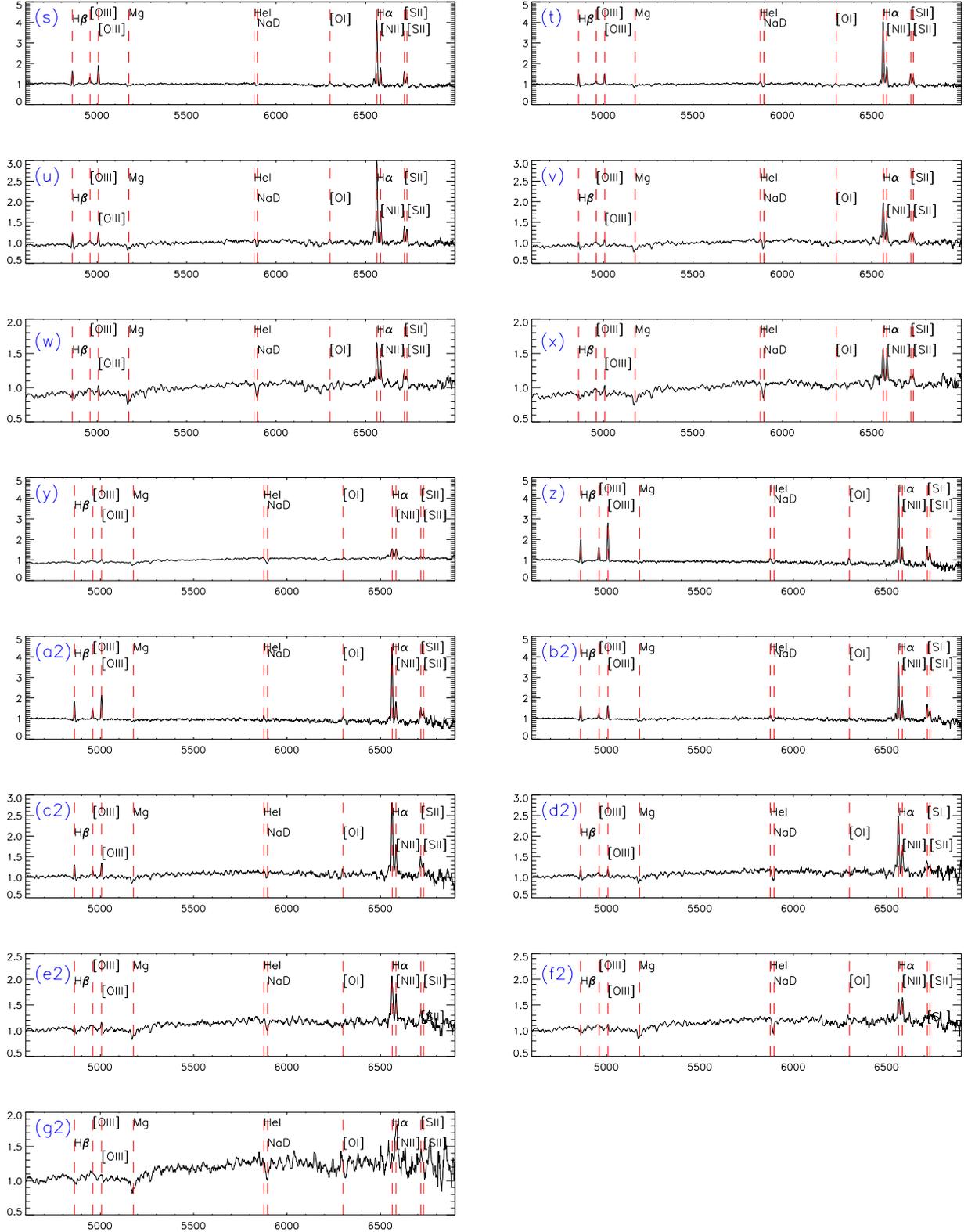}
 \caption {\label{stacked} Stacked zCOSMOS spectra at lower z. Bin 
$9.25\leq \log M_*/M_{\odot}<9.5$, $0.177 \leq z<0.242$ (panel a);
$9.5\leq \log M_*/M_{\odot}<9.75$, $0.177 \leq z<0.242$ (panel b); 
$9.75\leq \log M_*/M_{\odot}<10$, $0.177 \leq z<0.242$ (panel c);
$10\leq \log M_*/M_{\odot}<10.25$, $0.177 \leq z<0.242$ (panel d); 
$10.25\leq \log M_*/M_{\odot}<10.5$, $0.177 \leq z<0.242$ (panel e); $10.5\leq \log M_*/M_{\odot}<10.75$, $0.177 \leq z<0.242$ (panel f); $10.75\leq \log M_*/M_{\odot}<11$, $0.177 \leq z<0.242$ (panel g); $9\leq \log M_*/M_{\odot}<9.25$, $0.242 \leq z<0.306$ (panel h); $9.25\leq \log M_*/M_{\odot}<9.5$, $0.242 \leq z<0.306$ (panel i); $9.5\leq \log M_*/M_{\odot}<9.75$, $0.242 \leq z<0.306$ (panel j); $9.75\leq \log M_*/M_{\odot}<10$, $0.242 \leq z<0.306$ (panel k); $10\leq \log M_*/M_{\odot}<10.25$, $0.242 \leq z<0.306$ (panel l); $10.25\leq \log M_*/M_{\odot}<10.5$, $0.242 \leq z<0.306$ (panel m); $10.5\leq \log M_*/M_{\odot}<10.75$, $0.242 \leq z<0.306$ (panel n); $10.75\leq \log M_*/M_{\odot}<11$, $0.242 \leq z<0.306$ (panel o); $11\leq \log M_*/M_{\odot}<11.25$, $0.242 \leq z<0.306$ (panel p); $9.25\leq \log M_*/M_{\odot}<9.5$, $0.306 \leq z<0.371$ (panel q); $9.5\leq \log M_*/M_{\odot}<9.75$, $0.306 \leq z<0.371$ (panel r);
 $9.75\leq \log M_*/M_{\odot}<10$, $0.306 \leq z<0.371$ (panel s); $10\leq \log M_*/M_{\odot}<10.25$, $0.306 \leq z<0.371$ (panel t); $10.25\leq \log M_*/M_{\odot}<10.5$, $0.306 \leq z<0.371$ (panel u); $10.5\leq \log M_*/M_{\odot}<10.75$, $0.306 \leq z<0.371$ (panel v); $10.75\leq \log M_*/M_{\odot}<11$, $0.306 \leq z<0.371$ (panel w); $11\leq \log M_*/M_{\odot}<11.25$, $0.306 \leq z<0.371$ (panel x); $9.25\leq \log M_*/M_{\odot}<9.5$, $0.371 \leq z \leq 0.436$ (panel y); $9.5\leq \log M_*/M_{\odot}<9.75$, $0.371 \leq z \leq 0.436$ (panel z); $9.5\leq \log M_*/M_{\odot}<9.75$, $0.371 \leq z \leq 0.436$ (panel a2); $9.75\leq \log M_*/M_{\odot}<10$, $0.371 \leq z \leq 0.436$ (panel b2); $10\leq \log M_*/M_{\odot}<10.25$, $0.371 \leq z \leq 0.436$ (panel c2); $10.25\leq \log M_*/M_{\odot}<10.5$, $0.371 \leq z \leq 0.436$ (panel d2); $10.5\leq \log M_*/M_{\odot}<10.75$, $0.371 \leq z \leq 0.436$ (panel e2); $10.75\leq \log M_*/M_{\odot}<11$, $0.371 \leq z \leq 0.436$ (panel f2); $11\leq \log M_*/M_{\odot}<11.25$, $0.371 \leq z \leq 0.436$ (panel g2);}
\end{figure*}

\begin{figure*} 
  \centering
  \includegraphics[width=17cm,angle=0]{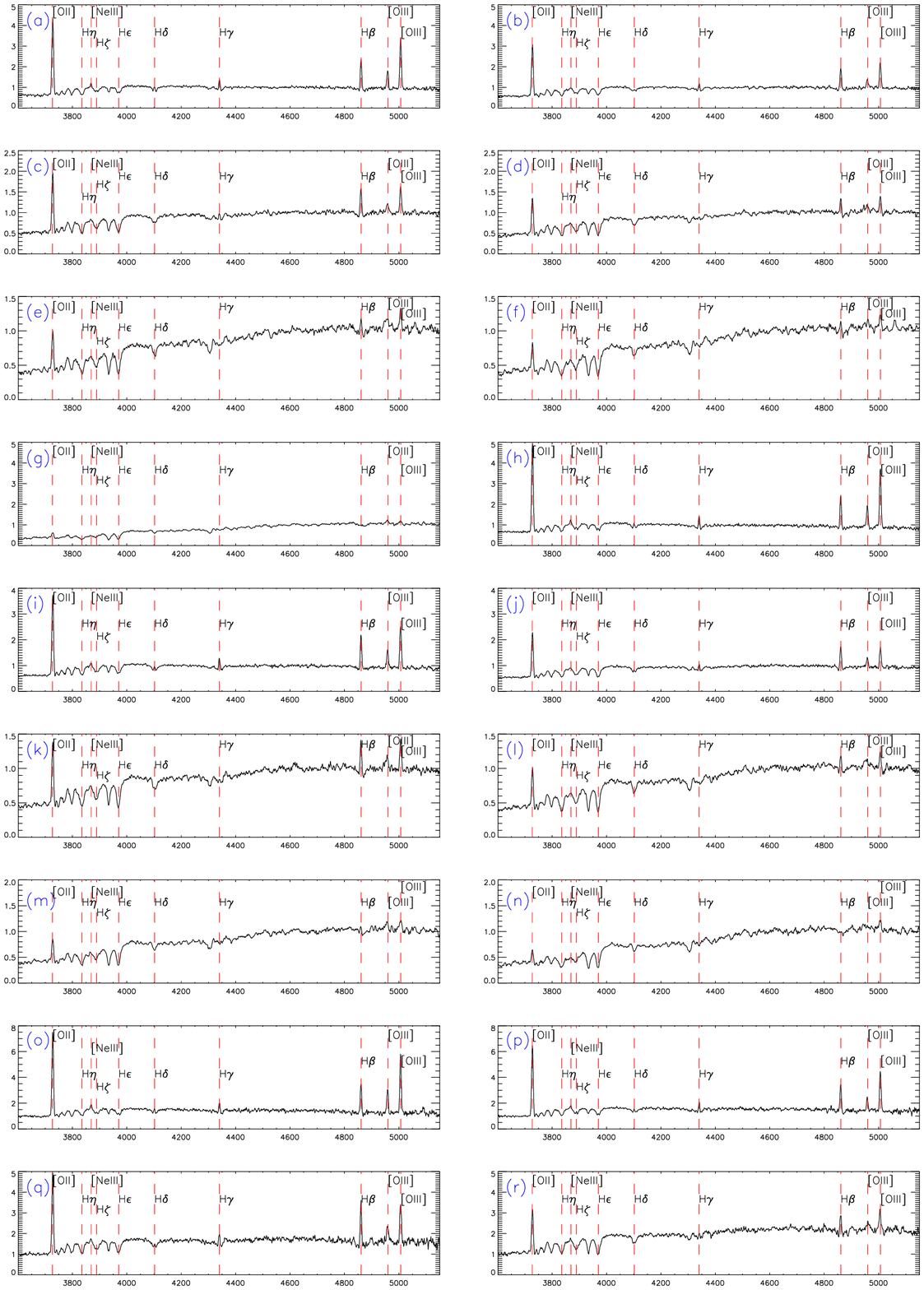}
\end{figure*}

\begin{figure*} 
  \centering
  \includegraphics[width=17cm,angle=0]{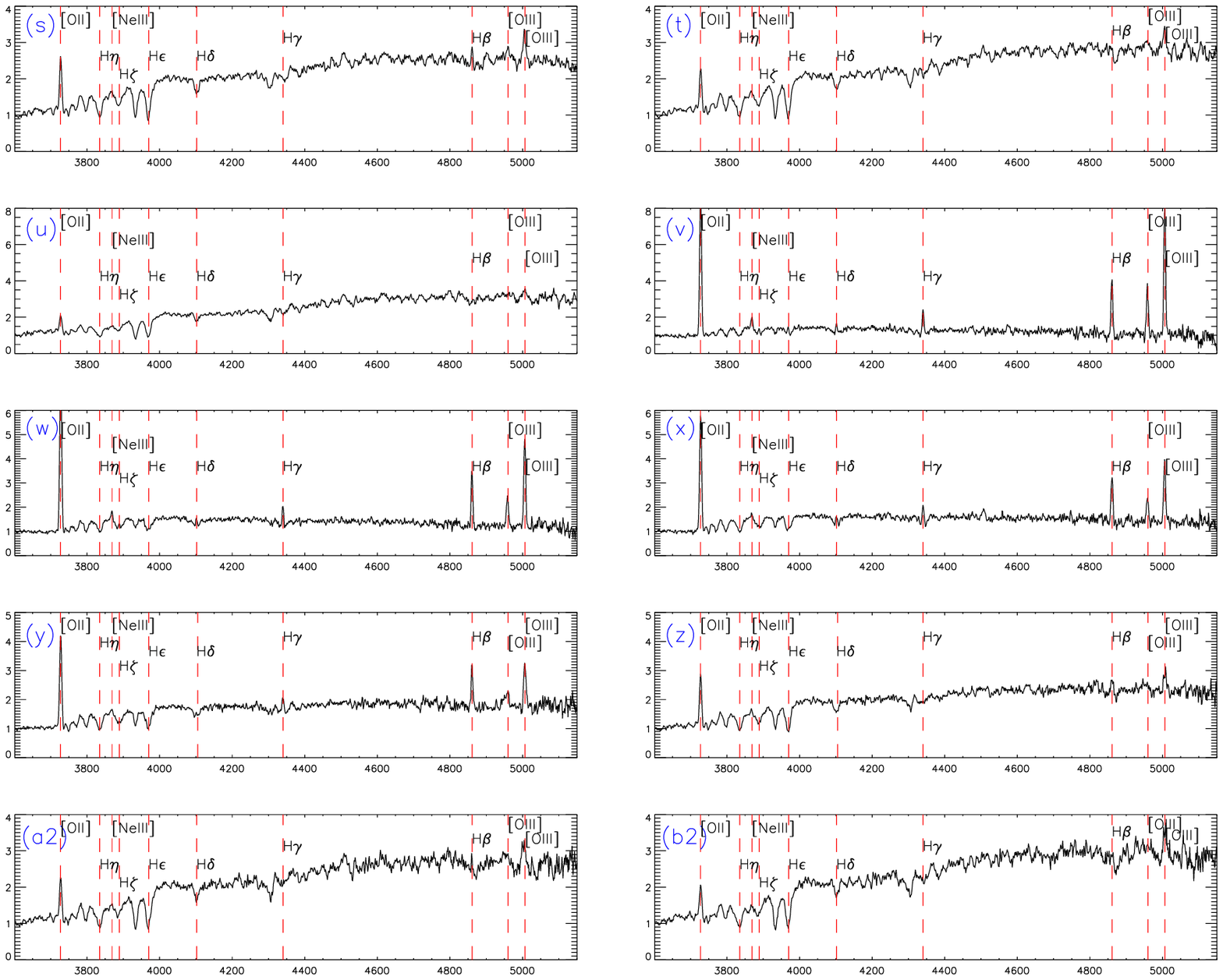}
 \caption {\label{stacked2} Stacked zCOSMOS spectra at higher z. Bin $9.5\leq \log M_*/M_{\odot}<9.75$, $0.548 \leq z<0.632$ (panel a); $9.75\leq \log M_*/M_{\odot}<10$, $0.548 \leq z<0.632$ (panel b); $10\leq \log M_*/M_{\odot}<10.25$, $0.548 \leq z<0.632$ (panel c); $10.25\leq \log M_*/M_{\odot}<10.5$, $0.548 \leq z<0.632$ (panel d); $10.5\leq \log M_*/M_{\odot}<10.75$, $0.548 \leq z<0.632$ (panel e); $10.75\leq \log M_*/M_{\odot}<11$, $0.548 \leq z<0.632$ (panel f); $11\leq \log M_*/M_{\odot}<11.25$, $0.548 \leq z<0.632$ (panel g); $9.5\leq \log M_*/M_{\odot}<9.75$, $0.632 \leq z<0.716$ (panel h); $9.75\leq \log M_*/M_{\odot}<10$, $0.632 \leq z<0.716$ (panel i); $10\leq \log M_*/M_{\odot}<10.25$, $0.632 \leq z<0.716$ (panel j); $10.25\leq \log M_*/M_{\odot}<10.5$, $0.632 \leq z<0.716$ (panel k); $10.5\leq \log M_*/M_{\odot}<10.75$, $0.632 \leq z<0.716$ (panel l); $10.75\leq \log M_*/M_{\odot}<11$, $0.632 \leq z<0.716$ (panel m); $11\leq \log M_*/M_{\odot}<11.25$, $0.632 \leq z<0.716$ (panel n);
 $9.5\leq \log M_*/M_{\odot}<9.75$, $0.716 \leq z<0.08$ (panel o); $9.75\leq \log M_*/M_{\odot}<10$, $0.716 \leq z<0.08$ (panel p); $10\leq \log M_*/M_{\odot}<10.25$, $0.716 \leq z<0.08$ (panel q); $10.25\leq \log M_*/M_{\odot}<10.5$, $0.716 \leq z<0.08$ (panel r); $10.5\leq \log M_*/M_{\odot}<10.75$, $0.716 \leq z<0.08$ (panel s); $10.75\leq \log M_*/M_{\odot}<11$, $0.716 \leq z<0.08$ (panel t); $11\leq \log M_*/M_{\odot}<11.25$, $0.716 \leq z<0.08$ (panel u);
 $9.5\leq \log M_*/M_{\odot}<9.75$, $0.8 \leq z \leq 0.884$ (panel v); $9.75\leq \log M_*/M_{\odot}<10$, $0.8 \leq z \leq 0.884$ (panel w); $10\leq \log M_*/M_{\odot}<10.25$, $0.8 \leq z \leq 0.884$ (panel x); $10.25\leq \log M_*/M_{\odot}<10.5$, $0.8 \leq z \leq 0.884$ (panel y); $10.5\leq \log M_*/M_{\odot}<10.75$, $0.8 \leq z \leq 0.884$ (panel z); $10.75\leq \log M_*/M_{\odot}<11$, $0.8 \leq z \leq 0.884$ (panel a2); $11\leq \log M_*/M_{\odot}<11.25$, $0.8 \leq z \leq 0.884$ (panel b2);}
\end{figure*}
For a fixed redshift, the stacked spectra - before stellar continuum subtraction - show a [OIII] line whose strength decreases with the progressive increase of the stellar mass (Fig. \ref{stacked}), especially if compared with the H$\beta$ line. The latter shows a weakening as well. The H$\alpha$ line is very strong at low masses. The intensity of the [NII] and H$\alpha$ decreases up to the point, at the highest mass bin, where these lines show the same intensity. In Fig.\ref{stacked2}, the [OIII] and H$\beta$ lines show the same trend as at lower redshifts. The [OII] line decreases with the increase of the mass, but it is always the strongest line in the high-redshift composite spectra.
\end{appendix}

\begin{appendix}
\section \newline
In this work we used the dust-corrected flux ratios for [OII]/H$\beta$ instead of the EW ratios originally used by \citet{Lamareille2010}.
  \begin{figure*} 
   \includegraphics[width=17cm,angle=0]{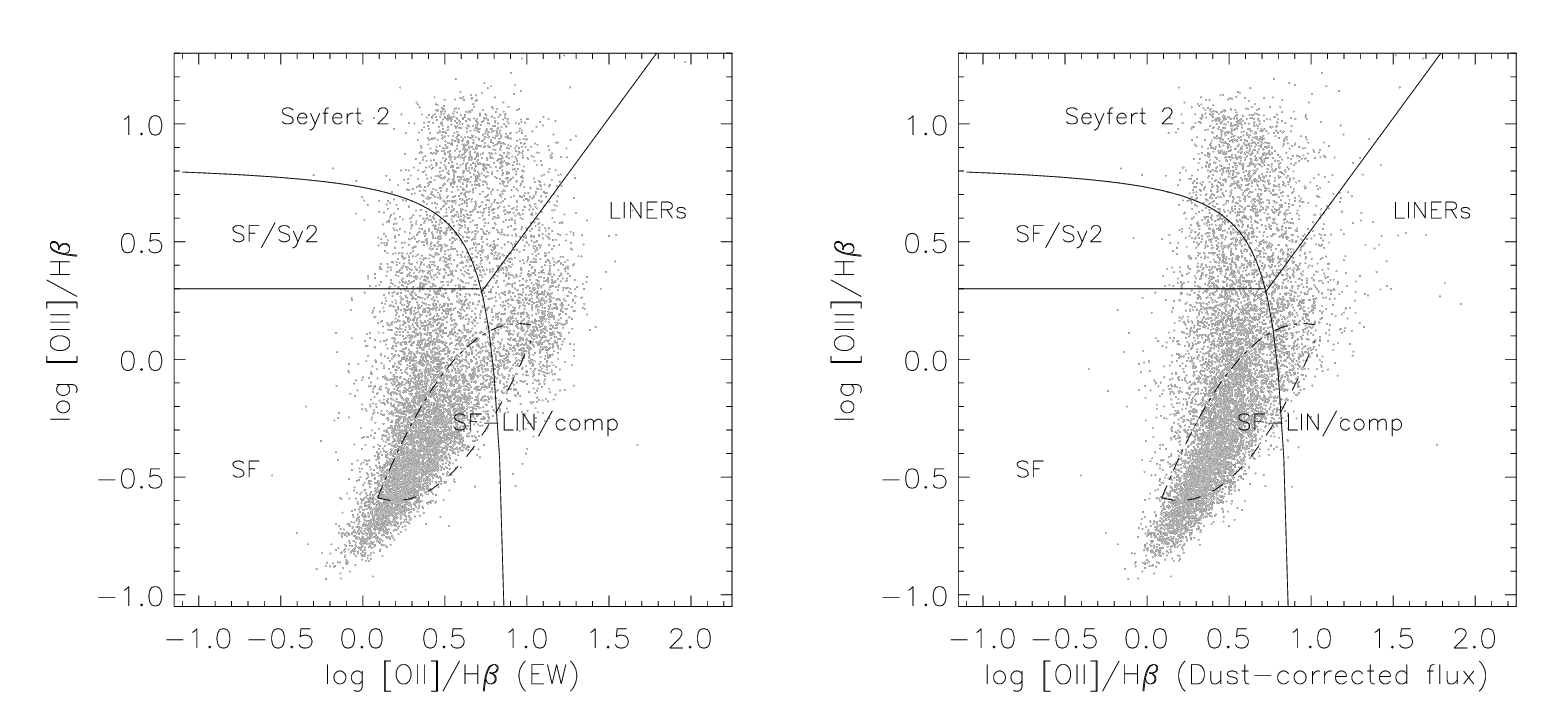}
   \caption{[OII] diagnostic diagram showing the consistency between EW measurements (left-panel) and dust-corrected fluxes (right panel). The solid demarcation lines are by \citet{Lamareille2010}. The gray distribution represents the SDSS galaxies from \citet{Vitale2012}.}
  \label{appendix}
  \end{figure*} 
This appendix (Fig.\ref{appendix}) shows that the dust-corrected fluxes correspond well enough to the EWs for a sample of SDSS galaxies \citep{Vitale2012}, especially in the regions occupied by our sample. Hence we can use the original demarcation lines from \citet{Lamareille2010}. However, the dust-corrected flux ratios span a slightly more restricted range of [OIII]/H$\beta$ values than the EW ratios.

\end{appendix}

\end{document}